\documentclass[11pt,graphicx,amsmath]{article}
\usepackage{amsmath}
\usepackage{graphicx}
\usepackage{epstopdf}
\usepackage{amsmath}
\usepackage{graphicx}
\usepackage{bm}
\usepackage{amssymb}
\usepackage{amsfonts}
\usepackage{mathtools}
\makeatletter

\newcommand{\Rmnum}[1]{\expandafter\@slowromancap\romannumeral #1@}
\makeatother
\usepackage{cite}
\usepackage{todonotes}
\usepackage{float}
\usepackage{subcaption}

\def\be{\begin{equation}}
\def\ee{\end{equation}}
\def\ba{\begin{eqnarray}}
\def\ea{\end{eqnarray}}
\def\nn{\nonumber}

\usepackage{amsmath}
\usepackage{graphicx}
\usepackage{bm}
\usepackage{xcolor}
\usepackage{amssymb}
\usepackage{amsfonts}
\usepackage{comment}
\usepackage{cite}
\usepackage{todonotes}
\usepackage{caption}
\usepackage{subcaption}
\usepackage{CJKutf8}
\usepackage{authblk}
\usepackage[utf8]{inputenc}

\def\be{\begin{equation}}
\def\ee{\end{equation}}
\def\ba{\begin{eqnarray}}
\def\ea{\end{eqnarray}}
\def\nn{\nonumber}

\def\bl#1\el{\begin{align}#1\end{align}}

\usepackage{amsmath}
\usepackage{graphicx}
\usepackage{bm}
\usepackage{color}
\usepackage{amssymb}
\usepackage{amsfonts}
\usepackage{comment}
\usepackage{cite}
\usepackage{todonotes}
\usepackage{appendix}

\allowdisplaybreaks
\sf
\large

\def\bl#1\el{\begin{align}#1\end{align}}

\title{Vacuum stress tensor of spinor fields in de Sitter space}
\author[a,b]{Xuan Ye\thanks{ yexuan@cuhk.edu.cn}}
\author[b]{Yang Zhang\thanks{ yzh@ustc.edu.cn}}
\affil[a]{\small School of Science and Engineering, The Chinese University of Hong Kong (Shenzhen), Longgang, Shenzhen, Guangdong, 518172, P.R. China}
\affil[b]{\small Department of  Astronomy,  Key Laboratory
for Researches in Galaxies and Cosmology,
 University of Science and Technology of China,  Hefei, Anhui, 230026,  China}

\begin{document}

\maketitle
\begin{abstract}
\small

We study the regularized vacuum stress tensor of spinor fields in de Sitter space and its possible contribution to the cosmological constant. Using adiabatic regularization, we show that the regularized stress tensor is maximally symmetric. It vanishes for the massless field, while for the massive field the vacuum energy density remains negative. The spinor vacuum therefore cannot by itself account for the observed positive cosmological constant, although it may contribute to the total vacuum energy.
We revisit the adiabatic expansion of the spinor mode functions and find that several arbitrary functions in the WKB expansion cannot be fixed by the conditions given in the literature. However, their dependence cancels in the adiabatic power spectrum and spectral stress tensor, so that the physical quantities are uniquely determined. We further show that, in a general flat Robertson-Walker spacetime, the 2nd order adiabatic regularization is sufficient to remove all ultraviolet divergences in both quantities.
Using point-splitting regularization, we derive the analytic correlation functions for both massless and massive spinor fields in de Sitter space that are valid at all scales, together with the adiabatic correlation functions. The regularized stress tensor obtained analytically in coordinate space agrees with those obtained from numerical integration of the adiabatically regularized spectra.
The 2nd order regularized correlation function and stress tensor of the massive field reduce smoothly to those of the massless field in the massless limit, whereas the 4th order regularized results lack this property.

\end{abstract}

\section{Introduction}

The accelerated expansion of the Universe is commonly described by a positive cosmological constant or by a dark energy with an equation of state close to $w=-1$ \cite{Riess1998,Perlmutter1999,Planck2018,Alam2021}. A cosmological constant gives a constant and maximally symmetric stress tensor and leads to de Sitter expansion. De Sitter space also provides a useful approximation to the inflationary stage of the early Universe. Despite its simple form and its success in cosmology, the physical origin of the cosmological constant remains unclear \cite{weinberg1989}. One possible contribution comes from the vacuum stress tensor of quantum fields.

Quantum fields  in curved spacetime
\cite{Lparker1,Lparker2,DeWitt1975,Birrell1982,Parker2009,Fulling1989,Hu2020}
have ultraviolet (UV) divergences in the stress tensor in the vacuum state. These vacuum contributions cannot simply be removed by normal ordering, since their finite part can have gravitational effects in curved spacetime and may contribute to the cosmological constant \cite{Feynman,Zeldovich1968}. A proper regularization is therefore required before the physical role of the vacuum stress tensor can be discussed. Two widely used methods are adiabatic regularization in momentum space \cite{ParkerFulling1974,FullingParkerHu1974,HuParker1978,BLHu1978,Birrell1978,Bunch1978,BunchParker1979,Bunch1980} and point-splitting regularization in coordinate space \cite{DeWitt1975,Christensen1976,Christensen1978,BunchDavies1978,pointsplittingyzw2022}. The former gives the regularized spectral quantities and their momentum integrals, while the latter provides a direct calculation of the regularized correlation function and stress tensor in coordinate space. In this paper, we use adiabatic regularization as the main method, constructing subtraction terms from the adiabatic modes. As a complementary approach, point-splitting regularization yields analytic results in coordinate space, which are consistent with those obtained from the adiabatic regularization.

In our previous work, we studied the vacuum regularization of scalar, vector, and tensor fields in cosmological spacetimes \cite{yangzhang20201,yangzhang20202,pointsplittingyzw2022,ZhangYe2022,YeZhang2024,XuanZhang2025,ZhangYeGW2025}. Following the minimal subtraction rule \cite{ParkerFulling1974}, we found that the required adiabatic order depends on the field, its coupling, and its components. For example, the 0th and 2nd orders are sufficient for conformally and minimally coupled scalar fields in de Sitter space, respectively \cite{yangzhang20201,pointsplittingyzw2022}. For the Stueckelberg field, different components require different subtraction orders \cite{XuanZhang2025}, while the stress tensor of gravitational waves in a flat Robertson-Walker (fRW) spacetime can be regularized at the 2nd order \cite{FordParker1977,ZhangYeGW2025}. In these studies, the regularized stress tensors of massive fields are maximally symmetric in de Sitter space, showing that quantum vacuum fields can in principle contribute to a cosmological constant.

In this paper, we extend this study to spinor fields. Systematic adiabatic expansions for spinor mode functions in a fRW spacetime were developed in Refs. \cite{ChimentoCossarini1990,Landete201314,Landete2013142,1Rio2014,Rio2017,Fernando2018,Ghosh2015,Rio2015}. We revisit this construction and find that several arbitrary functions in the WKB expansion cannot be uniquely fixed by the conditions used in the literature. We keep these functions throughout the calculation and show that their contributions cancel in the adiabatic power spectrum and spectral stress tensor. The corresponding physical quantities are therefore uniquely determined. We further show that the 2nd order adiabatic regularization is sufficient to remove all UV divergences in both quantities in a general fRW spacetime. 

We then study the vacuum stress tensor in de Sitter space. For the massless spinor field, the 2nd order regularized stress tensor and its trace vanish. For the massive field, the regularized stress tensor is maximally symmetric and its energy density remains negative. We also derive the analytic correlation functions for both massless and massive spinor fields that are valid at all scales, together with the adiabatic correlation functions. This allows us to perform the point-splitting regularization analytically. The regularized stress tensor agree with those obtained from numerical integration of the adiabatically regularized spectra. In addition, the 2nd order regularized correlation function and stress tensor of the massive field reduce smoothly to those of the massless field in the massless limit. The 4th order regularized results do not share this smooth limit. These results show that the massless spinor vacuum gives no contribution to the cosmological constant, while the massive spinor field, because of its negative vacuum energy density, cannot by itself account for the observed positive cosmological constant, although it may contribute to the total vacuum energy.

The paper is organized as follows. 
Sec. \ref{sec2} presents the exact mode functions for spinor fields in de Sitter space and the adiabatic mode functions in a general fRW spacetime. 
Sec. \ref{sec3} studies the adiabatic regularization of the power spectrum. 
Sec. \ref{sec4} applies the adiabatic regularization to the spectral stress tensor and gives the corresponding numerical results. 
Sec. \ref{sec5} gives the point-splitting regularization and
derives the analytic expressions for the regularized correlation functions and stress tensor. 
Sec. \ref{sec6} presents our conclusions and discussion. 
Appendix \ref{AAA} examines the WKB modes with the arbitrary functions up to the 4th order
and the treatment differs from Refs. \cite{Landete201314,Landete2013142,1Rio2014}.
Appendix \ref{BBB} shows that the arbitrary functions cancel out
in the adiabatic power spectrum and spectral stress tensor.
Appendix \ref{CCC} performs the integrations for the analytical correlation function
of the massive spin-$\frac12$ fields in de Sitter space.

We use natural units $c=\hbar=1$ throughout the paper.

\section{The adiabatic solutions of spin-1/2 field}\label{sec2}

The Lagrangian density of a spin-$\frac12$  field in  curved spacetime is given by \cite{Parker2009}
\bl
{\cal{L}}&=\sqrt{-g}\bar{\psi}(i\bar{\gamma}^\mu\nabla_\mu-m)\psi,
\label{Ldensity}
\el
where $\psi$ is the spinor field and $m$ is the mass. The spacetime dependent matrices $\bar\gamma^{\mu}(x)$ satisfy the anticommutation relation
$\{\bar{\gamma}^{\mu},\bar{\gamma}^{\nu}\}=2g^{\mu\nu}$ and are defined by
the tetrad fields {\small$V^{~\mu}_{a}$}
as {\small$\bar{\gamma}^{\mu}=V^{~\mu}_{a}\gamma^{a}$}, where $\gamma^{a}$ are the 4$\times$4 constant gamma matrices in Minkowski spacetime.
The covariant derivative acting on the spinor field is defined by
{\small$\nabla_{\mu}\equiv \partial_{\mu}-\Gamma_{\mu}$}, where
the spin connection is given by {\small $\Gamma_{\nu}=-\frac{1}{4}
\gamma_{a}\gamma_{b} V^{a\lambda}
V^{b}_{~\lambda;\nu}$}, with the semicolon denoting the covariant derivative acting on a tensor index.
In this work, we adopt the standard Dirac-Pauli representation, where the gamma matrices take the form
\bl
\gamma^0=\left(
  \begin{array}{ccc}
    I & 0 \\
    0 & -I \\
  \end{array}
\right),
~~~~~~~~~~~
\gamma^i=\left(
  \begin{array}{ccc}
    0 & \sigma^{i} \\
    -\sigma^{i} & 0 \\
  \end{array}
\right),\label{FSpacetimeDM}
\el
and $\sigma^i$ are the standard Pauli matrices.
The metric of the fRW
spacetime is
\bl
ds^2=dt^2-a(t)^2(dx^2+dy^2+dz^2),
\el
where $a$ is the scale factor and $t$ is the cosmic time with $\dot{a}=da/dt$.
In the fRW spacetime, the tetrad fields can be chosen as $
V_{\alpha}^{~\mu}=(1, a^{-1}, a^{-1}, a^{-1})$, which leads to the following
spin connection components \cite{Lparker1}
\bl
\Gamma^0=0,~~~~~\Gamma^i=\frac12\frac{\dot{a}}{a^2}\gamma^0\gamma^i,\label{affine}
\el
and the spacetime dependent gamma matrices are
\bl
\bar{\gamma}^{0}=\gamma^0,~~~~~
\bar{\gamma}^{i}=a^{-1}\gamma^{i}.\label{gamma}
\el
From the Lagrangian density \eqref{Ldensity}  follows the Dirac equation in curved spacetime
\bl
(i\bar{\gamma}^\mu\nabla_\mu-m)\psi=0.\label{EOM}
\el
By multiplying $(i \bar{\gamma}^\nu \nabla_\nu + m)$  on \eqref{EOM}
from the left and using the relation
$[\bar{\gamma}^\mu, \bar{\gamma}^\nu][\nabla_\mu, \nabla_\nu] \psi = R \psi$
with $R$ being the scalar curvature
(see (5.271) in Ref.\cite{Parker2009}), one has
\bl
(\nabla_\mu \nabla^\mu + m^2 + \tfrac{1}{4} R) \psi = 0 . \label{KGfunctionofspin-1/2}
\el
This formally  resembles the Klein-Gordon equation of the scalar field
with a coupling constant $\xi = 1/4$ \cite{yangzhang20201}.
However,  $\psi$ is not simply a set of four arbitrary scalar functions,
since $\psi$ has to satisfy the spinor equation \eqref{EOM}.
Using \eqref{affine} and \eqref{gamma},
the Dirac equation \eqref{EOM} can be written as \cite{Lparker2}
\bl
\Big(i \gamma^0 \partial_0 + \frac{3i}{2} \frac{\dot{a}}{a} \gamma^0
+ \frac{i}{a} \gamma^i \partial_i - m \Big) \psi = 0
. \label{exactlyEOM}
\el
The  field operator can be   expanded as  \cite{Bjorken1964,peskin}
\bl
\psi(x,t) = \int \frac{d^3 k}{(2\pi)^{3/2}}
\sum_{\lambda = \pm \frac{1}{2}} \bigl( A_{\vec{k},\lambda} u_{\vec{k},\lambda}(t) e^{i \vec{k} \cdot \vec{x}}
+ B_{\vec{k},\lambda}^\dag v_{\vec{k},\lambda}(t) e^{-i \vec{k} \cdot \vec{x}} \bigr), \label{Qutpsi}
\el
where $A_{\vec{k},\lambda}$ and $B_{\vec{k},\lambda}^\dag$
are the annihilation and creation operators
respectively for electrons and positrons with helicity $\lambda$
and  momentum $\vec{k}$,
and $u_{\vec{k},\lambda}$ and $v_{\vec{k},\lambda}$ are the mode spinors,
the anticommutation relations for these operators are
\bl
\{ A_{\vec{k},\lambda}, A_{\vec{k}',\lambda'}^\dag \}
= \{ B_{\vec{k},\lambda}, B_{\vec{k}',\lambda'}^\dag \}
= \delta_{\lambda\lambda'} \delta^{(3)}(\vec{k} - \vec{k}')
   . \label{anticommutation}
\el
Plugging \eqref{Qutpsi} into  \eqref{exactlyEOM} yields
\bl
i \gamma^0 \dot{u}_{\vec{k},\lambda} + \frac{3i}{2} \frac{\dot{a}}{a} \gamma^0 u_{\vec{k},\lambda}
- \frac{1}{a} \gamma^i k_i u_{\vec{k},\lambda} - m u_{\vec{k},\lambda} = 0, \label{uk}
\el
the spinor $v_{\vec{k},\lambda}$ can be obtained by charge conjugation,
$v_{\vec{k},\lambda} = -i \gamma^2 u_{\vec{k},\lambda}^*$ \cite{peskin}.
The spinor $u_{\vec{k},\lambda}$ can be expressed
in terms of the two-component spinors $\xi_{\lambda, \vec{k}}$
  as the following \cite{Landete201314,Landete2013142,1Rio2014,Rio2017}
\bl
u_{\vec{k},\lambda}(t)= \frac{1}{ a^{\frac32}}\left(
  \begin{array}{ccc}
   h_k^{\Rmnum1}(t)\xi_{\lambda,\vec{k}}  \\
   h_k^{\Rmnum2}(t)\frac{\sigma^ik_i}{k}\xi_{\lambda,\vec{k}}       \\
  \end{array}
\right),\label{uklambda}
\el
where
\bl
\xi_{\frac12,\vec{k}}=\left(
  \begin{array}{ccc}
  \cos(\frac{\theta_k}{2}) e^{-i\phi_k} \\
       \sin(\frac{\theta_k}{2}) \\
  \end{array}
  \right),~~~
\xi_{-\frac12,\vec{k}}=\left(
  \begin{array}{ccc}
  -\sin(\frac{\theta_k}{2}) e^{-i\phi_k} \\
       \cos(\frac{\theta_k}{2}) \\
  \end{array}
  \right),\label{xi}
\el
with $\theta_k$ and $\phi_k$ being the polar and azimuthal angles of $\vec{k}$
in momentum space.  The spinors $\xi_{\lambda, \vec{k}}$ satisfy
the eigenvalue equation
$\frac{\sigma^i k_i}{2k} \xi_{\lambda, \vec{k}} = \lambda \xi_{\lambda, \vec{k}}$
with the normalization
$\xi_{\lambda, \vec{k}}^\dag  ~\xi_{\lambda', \vec{k}} = \delta_{\lambda \lambda'}$.
Using the equal-time anticommutation relation
$\{\psi_a(x,t), \pi_b(x',t)\} = i \delta(x - x') \delta_{ab}$,
where the canonical momentum is defined by
$\pi \equiv \frac{\partial \mathcal{L}}{\partial (\partial_0 \psi)} = \sqrt{-g} \, i \psi^{\dagger}$,
and with $(a,b)$ denoting spinor indices, together with \eqref{anticommutation},
one finds the normalization condition
\bl
|h_k^{\Rmnum{1}}|^2 + |h_k^{\Rmnum{2}}|^2 = 1. \label{normalcon}
\el
Plugging \eqref{uklambda} into \eqref{uk} yields two coupled first order differential equations
\begin{subequations}
\bl
h_k^{\Rmnum1}(t)
&= i \frac{a}{k} (\partial_0 - i m) \, h_k^{\Rmnum2}(t),
\label{oneorderequ1} \\
h_k^{\Rmnum2}(t)
&= i \frac{a}{k} (\partial_0 + i m) \, h_k^{\Rmnum1}(t) .
\label{oneorderequ2}
\el
\end{subequations}
The functions $ h_k^{\Rmnum1}$ and $h_k^{\Rmnum2}$ contain  the variable $t$,
and also depend on the  parameter $m$.
Eqs.\eqref{oneorderequ1} and \eqref{oneorderequ2} imply the following relation
\bl
h_k^{\Rmnum1}(t; -m) = h_k^{\Rmnum2}(t; m). \label{300}
\el
In de Sitter space, the scale factor is given by
\bl
a(t) = e^{H t}, \label{scalefactor}
\el
where $H$ is the Hubble parameter.
Eqs.~\eqref{oneorderequ1} and \eqref{oneorderequ2} can be rewritten as two decoupled second order differential equations
\begin{subequations}
\bl
&z^{2} \frac{\partial^{2} h_k^{\Rmnum1}(z)}{\partial z^{2}}
+ z \frac{\partial h_k^{\Rmnum1}(z)}{\partial z}
+ \left( z^{2} - (-i \mu + \tfrac{1}{2})^{2} \right) h_k^{\Rmnum1}(z) = 0, \label{equationH1} \\
&z^{2} \frac{\partial^{2} h_k^{\Rmnum2}(z)}{\partial z^{2}}
+ z \frac{\partial h_k^{\Rmnum2}(z)}{\partial z}
+ \left( z^{2} - (-i \mu - \tfrac{1}{2})^{2} \right) h_k^{\Rmnum2}(z) = 0, \label{equationH2}
\el
\end{subequations}
where  the dimensionless variable $z \equiv k / (a H)$ and
the dimensionless parameter $\mu \equiv  m / H  $.
Eqs. \eqref{equationH1} and \eqref{equationH2} can
be alternatively derived by substituting \eqref{Qutpsi} and \eqref{uklambda}
into \eqref{KGfunctionofspin-1/2}.

The positive frequency solutions of \eqref{equationH1} and \eqref{equationH2} are
\cite{Landete2013142,Olver2010}
\begin{subequations}
\bl
h_k^{\Rmnum1}(z) &=i\frac{\sqrt{\pi z}}{2}
e^{\frac{\pi\mu}{2}}H^{(1)}_{-i\mu+\frac12}(z),\label{exactlywavefunction1}
\\
h_k^{\Rmnum2}(z) &=\frac{\sqrt{\pi z}}{2}
      e^{\frac{\pi\mu}{2}}H^{(1)}_{-i\mu-\frac12}(z),\label{exactlywavefunction2}
\el
\end{subequations}
where $H^{(1)}_{\nu}(z)$  is the Hankel function of the first kind.
The modes $h_k^{\Rmnum1}$ and $h_k^{\Rmnum2}$
satisfy the relation \eqref{300} and the normalization condition \eqref{normalcon}.
In the massless limit $\mu \rightarrow 0$, the exact modes
\eqref{exactlywavefunction1} and  \eqref{exactlywavefunction2} reduce to
\begin{subequations}
\bl
\lim_{\mu\rightarrow0} h_k^{\Rmnum1}(z)&=\frac{1}{\sqrt{2}}e^{iz},\label{exactlywavefunctionmasless1}
\\
\lim_{\mu\rightarrow0} h_k^{\Rmnum2}(z)&=\frac{1}{\sqrt{2}}e^{iz}.\label{exactlywavefunctionmasless2}
\el
\end{subequations}
The solutions \eqref{exactlywavefunction1} - \eqref{exactlywavefunctionmasless2}
will be used to compute the unregularized  vacuum correlation function and stress tensor.

To analyze the  vacuum UV divergences,
we next examine the high frequency behavior of $h^{\Rmnum1}_k$ and  $h^{\Rmnum2}_k$.
The WKB modes will be used in adiabatic regularization \cite{ParkerFulling1974}
 because they approximate the exact modes at high $k$ adequately
 and respect the conservation to each adiabatic order.
Assuming the $n$th  order  WKB approximations  for $h^{\Rmnum1}_k$ and  $h^{\Rmnum2}_k$
have the following  form   \cite{Landete201314,Landete2013142,1Rio2014}
\begin{subequations}
\bl
g^{I(n)}_k(t)&=\sqrt{\frac{\omega+m}{2\omega}}e^{-i\int^{t}\Omega(t')dt'}F(t)
,\label{45}
\\
g^{II(n)}_k(t)&=\sqrt{\frac{\omega-m}{2\omega}}e^{-i\int^{t}\Omega(t')dt'}G(t)
,\label{46}
\el
\end{subequations}
where $\omega =  \sqrt{ k^2/a^2 +m^2}$, and the functions
\begin{subequations}
\bl
& \Omega(t)=\sum_{n=0} (\omega  +\omega^{(1)} + ... + \omega^{(n)}    ),
\label{27}
\\
& F(t)=\sum_{n=0} (F^{(0)} + F^{(1)}  + ... + F^{(n)}   ),
\label{28}
\\
& G(t)=\sum_{n=0} (G^{(0)} + G^{(1)}  + ... + G^{(n)}   ) .
\label{47}
\el
\end{subequations}
At the 0th order,  $\Omega^{(0)}= \omega$, $G^{(0)}=F^{(0)}=1$.
At each order the WKB modes  $g^{I(n)}_k$ and $g^{II(n)}_k$ satisfy
the equations similar to   \eqref{oneorderequ1} and \eqref{oneorderequ2},
and  satisfy  the normalization condition
\bl
|g^{I(n)}|^2+|g^{II(n)}|^2=1,\label{normalizationforg}
\el
 and the relation
\bl
g^{I(n)}(t; -m)=g^{II(n)}(t; m) ,
 \label{relg1g2}
\el
in analogy to \eqref{normalcon} and \eqref{300}.
Plugging  \eqref{27}  \eqref{28} \eqref{47} into \eqref{relg1g2} leads to the following relations
\begin{subequations}
\bl
F^{(n)}(t; -m)&=G^{(n)}(t; m),\label{relFGn1}
\\
\omega^{(n)}(t; -m)&=\omega^{(n)}(t; m).\label{relFGn2}
\el
\end{subequations}
As shown in Appendix \ref{AAA},   in determining   $\Omega^{(n)}$, $F^{(n)}$ and $G^{(n)}$,
some arbitrary functions appear.
In Appendix \ref{BBB},
we show that all these arbitrary functions
cancel in the power spectrum and spectral stress tensor,
so they can be set to zero without affecting the physical results.

In the massless limit,  \eqref{45} and \eqref{46} become
\begin{subequations}
\begin{align}
g_k^{I(0)} = g_k^{I(2)} = g_k^{I(4)} &= \frac{1}{\sqrt{2}} e^{-i \int^{t} \frac{k}{e^{H t'}} dt'} = \frac{1}{\sqrt{2}} e^{i z}, \label{fhuier1} \\
g_k^{II(0)} = g_k^{II(2)} = g_k^{II(4)} &= \frac{1}{\sqrt{2}} e^{-i \int^{t} \frac{k}{e^{H t'}} dt'} = \frac{1}{\sqrt{2}} e^{i z}, \label{fhuier2}
\end{align}
\end{subequations}
ie,  the WKB modes of all orders are equal to
 the exact \eqref{exactlywavefunctionmasless1} and \eqref{exactlywavefunctionmasless2}.
This is an important property of  the massless WKB modes.

\section{Adiabatic regularization of power spectrum   }\label{sec3}

Now  we study the  power spectrum of the spin-$\frac12$ field  $\psi$,
and examine its UV divergences.
Given  $\psi$ and  the vacuum state $|0\rangle $ defined by
\bl
A_{\vec{k},\lambda}|0\rangle=
B_{\vec{k},\lambda}|0\rangle=0 ,
\label{definationofvaccum}
\el
one considers  the vacuum expectation value  as the following
\bl
\langle0|\bar\psi(x)\psi(x)|0\rangle .
\label{selfcor}
\el
This is a scalar,
and referred to as the auto-correlation function of the field  $\psi$.
As shall be seen later,  the vacuum stress tensor is related to the auto-correlation function
in de Sitter space.

Using \eqref{Qutpsi} into \eqref{selfcor}, we get
\bl
\langle0|\bar\psi(x)\psi(x)|0\rangle
=\int \Delta_k^{2} \frac{d k}{k},
\label{barpsipsi}
\el
where  the vacuum  power spectrum is
\bl
~~~ \Delta_k^{2}
\equiv -\frac{k^3}{a^3\pi^2}(|h^{\Rmnum1}_k|^2-|h^{\Rmnum2}_k|^2),
\label{400}
\el
with an overall minus sign.
In de Sitter space, using the exact  modes \eqref{exactlywavefunction1} and \eqref{exactlywavefunction2},
the power spectrum is
\bl
\Delta^{2}_{k}
&=-\frac{ H^3}{\pi^2}z^3 \Big( |\frac{\sqrt{\pi z}}{2}
e^{\frac{\pi\mu}{2}} H^{(1)}_{-i\mu+\frac12}(z) |^2
-|\frac{\sqrt{\pi z}}{2}e^{\frac{\pi\mu}{2}}H^{(1)}_{-i\mu-\frac12}(z)|^2 \Big) .
   \label{unregde}
\el
\begin{figure}[htb]
  \centering
  \includegraphics[width=0.5\linewidth]{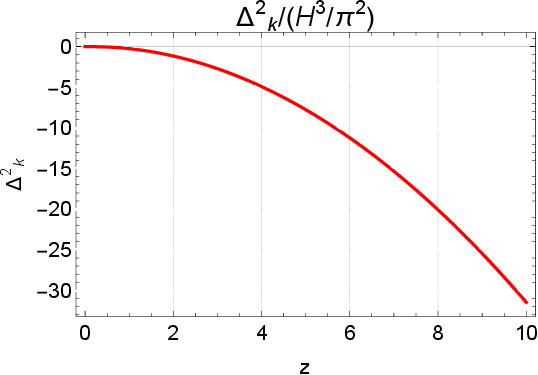}
  \caption{Unregularized power spectrum $\Delta^{2}_{k}$.
   The parameter $\mu^2 = \frac{m^2}{H^2} = 0.1$ is taken.}
  \label{fig1}
\end{figure}
Fig.\ref{fig1} shows that  $\Delta^{2}_{k} $ is negative and proportional to $ -k^2$ at high $k$.
(For illustration
the parameter $\mu^2  = 0.1$ will be taken in all the figures except for Fig. \ref{fig6}.)
This $\Delta^{2}_{k}$ will yield a UV divergent  auto-correlation
of \eqref{barpsipsi} at its upper limit of integration,
and in this sense the  power spectrum is said to be UV divergent.
A negative vacuum power spectrum is
due to the anticommutation relations \eqref{anticommutation} for the spin-$\frac12$  field operators.
This is in contrast to the scalar and vector fields
that have a positive vacuum  power spectrum.

The  high-z expansion of  \eqref{unregde} is
\bl
 &\lim_{z\rightarrow\infty}\Delta^{2}_{k} \simeq
   -\frac{ H^3}{\pi^2}z^3  \Big(\frac{\mu}{z}   -\frac{\mu \left(1+\mu^2\right)}{2 z^3}
      +\frac{3 \mu \left(4+5 \mu
        ^2+\mu^4\right)}{8 z^5}-\frac{5 \mu
  \left(1+\mu ^2\right) \left(4+\mu^2\right) \left(9+\mu^2\right)}{16 z^7}\nn
       \\
&+\frac{35 \mu \left(1+\mu^2\right) \left(4+\mu^2\right) \left(9+\mu^2\right)
    \left(16+\mu^2\right)}{128 z^9} -\frac{63 \mu \left(1+\mu^2\right)
    \left(4+\mu^2\right) \left(9+\mu^2\right) \left(16+\mu^2\right)
   \left(25+\mu^2\right)}{256 z^{11}}\nn
  \\
&~~~~~~~~~~+\frac{231 \mu  \left(1+\mu ^2\right) \left(4+\mu ^2\right)
 \left(9+\mu ^2\right) \left(16+\mu ^2\right)
  \left(25+\mu ^2\right) \left(36+\mu ^2\right)}{1024 z^{13}}
\Big),\label{highktyperwoexactly}
\el
where the first  two leading terms
\[
 -\frac{ H^3}{\pi^2}z^3  \Big( \frac{\mu}{z}   -\frac{\mu \left(1+\mu^2\right)}{2 z^3} \Big)
\]
are respectively  quadratic and logarithmic divergent.
The low-$z$  expansion  of  \eqref{unregde}  is
\bl
&\lim_{z\rightarrow0}\Delta^{2}_{k}
  \simeq  -\frac{ H^3}{\pi^2}\tanh(\pi  \mu)z^3 \leq 0,
\el
which is infrared convergent and negative.

Now we are to remove the UV divergences in the power spectrum of the spin-$\frac12$ field
by the adiabatic regularization.
For that purpose,
firstly we use  the WKB approximate modes \eqref{45} and \eqref{46}
to construct the adiabatic power spectra  $\Delta^{2(n)}_{k, ad}$ of $n$th order.
The  adiabatic power spectra for $n=0,2,4$
are  listed  in \eqref{B1}, \eqref{B3}, and \eqref{B6}.
(See Appendix \ref{BBB} for details.)
Then,  subtracting  $\Delta^{2(n)}_{k, ad}$ from the unregularized spectrum $\Delta^{2}_{k}$,
we get  the regularized power spectra   as the following
\bl
\Delta_{k~reg}^{2(n)}&\equiv\Delta_{k}^{2}-\Delta_{k~ad}^{2(n)},
~~ n=0,2,4,... .
 \label{reguidwelkn}
\el
 In de Sitter space, the adiabatic subtraction terms are explicitly
\begin{subequations}
\bl
\Delta^{2(0)}_{k~ad} &= - \frac{ H^3}{\pi^2}z^3 \big( \frac{\mu}{\bar\omega}  \big), \label{powerspectrumA0th}
\\
\Delta^{2(2)}_{k~ad}&
=-\frac{ H^3}{\pi^2}z^3\big(\frac{\mu}{\bar\omega}-\frac{\mu}{2 \bar\omega^3}
+\frac{9 \mu^3}{8 \bar\omega^5}-\frac{5\mu^5}{8 \bar\omega^7}  \big)
,\label{powerspectrumA2nd}
\\
\Delta^{2(4)}_{k~ad}&
=-\frac{H^3}{\pi^2}z^3\big(\frac{\mu}{\bar\omega}-\frac{\mu}{2 \bar\omega^3}
+\frac{3 \mu (4+3 \mu^2)}{8 \bar\omega^5}-\frac{5 \mu^3 (37+2 \mu^2)}{16 \bar\omega^{7}}
+\frac{3535 \mu^5}{128\bar\omega^{9}}-\frac{1701 \mu^7}{64 \bar\omega^{11}}
+\frac{1155 \mu^9}{128 \bar{\omega}^{13}}\big)
  ,\label{powerspectrumA4thh}
\el
\end{subequations}
where  $\bar{\omega}\equiv \omega/H= (z^2+\mu^2)^{1/2}$.
The high-$z$ limit of \eqref{powerspectrumA0th}, \eqref{powerspectrumA2nd}, and \eqref{powerspectrumA4thh}
are
\begin{subequations}
\bl
\lim_{z\rightarrow\infty}\Delta^{2(0)}_{k~ad}&\simeq-\frac{ H^3}{\pi^2}z^3
\big( \frac{\mu}{z} -\frac{\mu ^3}{2z^3}+\frac{3 \mu ^5}{8 z^5}-\frac{5 \mu ^7}
{16 z^7}+\frac{35 \mu ^9}{128 z^9}-\frac{63 \mu ^{11}}{256 z^{11}}+\frac{231
\mu ^{13}}{1024 z^{13}}  \big),\label{highzpowerspectrumA0th}
\\
\lim_{z\rightarrow\infty}\Delta^{2(2)}_{k~ad}&\simeq-\frac{ H^3}{\pi^2}z^3\Big(\frac{ \mu}{z} -\frac{\mu  \left(1+\mu ^2\right)}{2z^3} +\frac{3 \mu ^3 \left(5+\mu ^2\right)}{8 z^5}-\frac{5 \mu ^5 \left(14+\mu ^2\right)}{16 z^7}
\nn
\\
& +\frac{35 \mu ^7 \left(30+\mu ^2\right)}{128 z^9}
 -\frac{63 \mu ^9 \left(55+\mu ^2\right)}{256 z^{11}}+\frac{231 \mu ^{11}
\left(91+\mu ^2\right)}{1024 z^{13}}\Big),\label{highzpowerspectrumA2nd}
\\
\lim_{z\rightarrow\infty}\Delta^{2(4)}_{k~ad}&
\simeq -\frac{H^3}{\pi^2}z^3 \big( \frac{\mu}{z} -\frac{\mu\left(1+\mu ^2\right)}{2z^3}  +\frac{3 \mu  \left(4+5 \mu ^2+\mu ^4\right)}{8 z^5}
\nn \\
& -\frac{5 \mu ^3 \left(7+\mu ^2\right)^2}{16 z^7}+\frac{35 \mu ^5 \left(273+30 \mu ^2 +\mu ^4\right)}{128 z^9}
\nn
\\
&  -\frac{63 \mu ^7 \left(1023+55 \mu ^2
+\mu ^4\right)}{256 z^{11}}+\frac{231 \mu ^9 \left(3003+91 \mu ^2+\mu ^4\right)}{
  1024 z^{13}}\big).\label{A4th}
\el
\end{subequations}
Clearly,  the first two leading terms of the 0th order   \eqref{highzpowerspectrumA0th}
do not cancel all the divergences in the  unregularized  $\Delta_k^2$ of  \eqref{highktyperwoexactly},
and there still remains a logarithmic divergence $\frac{H^3}{\pi^2} \frac{\mu}{2}$.
The first two leading terms of the 2nd order  \eqref{highzpowerspectrumA2nd}
successfully cancel  all the divergences in $\Delta_k^2$  of \eqref{highktyperwoexactly},
yielding a UV convergent regularized power spectrum  $\Delta_{k\,reg}^{2(2)}$.
According to the minimal subtraction rule \cite{ParkerFulling1974},
the 2nd order regularization is sufficient.
The 2nd order regularized power spectrum at high  $z$  is
{\small
\bl
\lim_{z \to \infty} \Delta^{2(2)}_{k\,reg} \simeq
& -\frac{H^3}{\pi^2} z^3 \Big(
\frac{3 \mu}{2 z^5} - \frac{5 \mu (36 + 49 \mu^2)}{16 z^7} + \frac{35 \mu (576 + 820 \mu^2 + 273 \mu^4)}{128 z^9}
\nn \\
& - \frac{63 \mu \left(14400 + 11 \mu^2 (1916 + 695 \mu^2 + 93 \mu^4)\right)}{256 z^{11}}
\nn \\
& + \frac{231 \mu \left(518400 + 13 \mu^2 (59472 + 22792 \mu^2 + 3421 \mu^4 + 231 \mu^6)\right)}{1024 z^{13}}
\Big),
\el
}
where the leading term,
\bl
-\frac{H^3}{\pi^2} z^3 \left(\frac{3 \mu}{2 z^5}\right)  \propto - \frac{1}{z^2},
\el
is UV convergent and remains negative at high $z$.
So the 2nd order adiabatic regularization  not only removes all UV divergences,
but also preserves the negative sign of the  vacuum power spectrum at high $z$.
In literature
the 2nd order adiabatic regularization was first applied
upon the power spectrum of a minimally coupling massive scalar field Ref. \cite{Parker2007}.

The 4th order subtraction term \eqref{A4th} would subtract more than necessary
and result in an improper regularized power spectrum.
Let us examine the 4th order  regularized power spectrum  at  high $z$
\bl
\lim_{z \rightarrow \infty} \Delta^{2(4)}_{k\,reg} \simeq
& -\frac{H^3}{\pi^2} z^3 \Big(
-\frac{45 \mu}{4 z^7} + \frac{35 \mu (144 + 205 \mu^2)}{32 z^9}
 - \frac{63 \mu (14400 + 21076 \mu^2 + 7645 \mu^4)}{256 z^{11}}
 \nn \\
& + \frac{231 \mu (518400 + 773136 \mu^2 + 296296 \mu^4 + 44473 \mu^6)}{1024 z^{13}}
\Big),
\label{delt4reg}
\el
where the leading term,
\bl
-\frac{H^3}{\pi^2} z^3 \left(-\frac{45 \mu}{4 z^7}\right) \propto \frac{1}{z^4} > 0,
\el
is over-convergent and  positive.
This is because  the 4th order regularization subtracts too much,
so that  the convergent term $\propto - 1/z^2$ of $\Delta_k^2$ has been   subtracted.

\begin{figure}[htb]
  \centering
        {%
          \includegraphics[width = .45\linewidth]{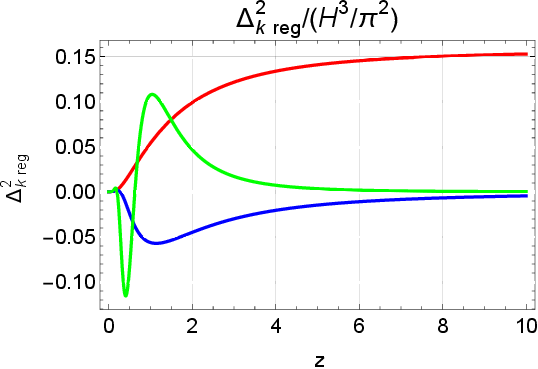}
  }
       {%
          \includegraphics[width = .45\linewidth]{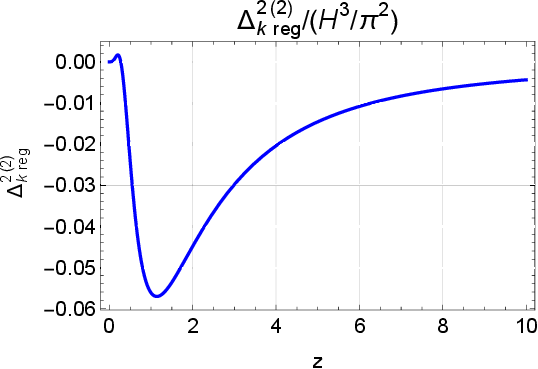}
      }
  \caption{
  (a) Regularized power spectrum:
  the 0th order $\Delta_{k\,reg}^{2(0)}$ (red),
  the 2nd order $\Delta_{k\,reg}^{2(2)}$ (blue),
  the 4th order $\Delta_{k\,reg}^{2(4)}$ (green).
  (b) The  enlarged 2nd order    $\Delta_{k~reg}^{2(2)}$.
  }
  \label{Powerspsp}
  \end{figure}

We plot the three regularized power spectra,
$\Delta_{k\,reg}^{2(0)}$, $\Delta_{k\,reg}^{2(2)}$, and
$\Delta_{k\,reg}^{2(4)}$ in Fig.~\ref{Powerspsp} (a)
with the parameter  $\mu^2  = 0.1$.
The 0th order regularized {\small$\Delta_{k\,reg}^{2(0)}$} (red) is UV logarithmically divergent  and positive.
The 4th order regularized {\small$\Delta_{k\,reg}^{2(4)}$} (green)
is over-convergent and positive at high $z$, showing an irregular infrared behavior.
The 2nd order regularized {\small$\Delta_{k\,reg}^{2(2)}$} (blue)
is UV  convergent and negative.
Fig.~\ref{Powerspsp} (b) shows an enlarged view of {\small$\Delta_{k\,reg}^{2(2)}$},
which at convergent  and   becomes positive at very small $z$,
\bl
\lim_{z \rightarrow 0} \Delta_{k\,reg}^{2(2)}
\simeq \frac{H^3}{\pi^2} z^3 (1- \tanh(\pi \mu)) .
\el
This infrared distortion is caused by the inaccuracy of WKB modes at small $k$
under the adiabatic regularization.
This issue has been addressed in the schemes of the inside-horizon regularization \cite{YangWang2018}
and the energy-dependent regularization \cite{FerreiroTorrenti}.

The massless limit ($\mu  \rightarrow 0$)  of   the unregularized power spectrum  \eqref{unregde} vanishes
\bl
\lim_{\mu \to 0} \Delta_k^2 = -\frac{H^3}{\pi^2} z^3 \times 0 = 0.
\label{pwms}
\el
The massless limits of  $\Delta_{k\,ad}^{2(2)}$  in  \eqref{powerspectrumA2nd} is vanishing
\bl
\lim_{\mu \rightarrow 0} \Delta^{2(2)}_{k\,ad}
  = 0,
 \label{2pw}
\el
so that the  massless limit of the 2nd order regularized spectrum is also vanishing,
\bl
\lim_{\mu \rightarrow 0} \Delta^{2(2)}_{k\,reg}
  = 0.
 \label{reglpower}
\el
From \eqref{powerspectrumA2nd} and \eqref{powerspectrumA4thh}
it is seen that the massless limits
\eqref{2pw} and \eqref{reglpower} are valid  for all the orders ($n=0,2,4, ...$).
If one starts with the massless field,
one also obtains  \eqref{2pw} and  \eqref{reglpower}.
Thus, under the adiabatic regularization,
the regularized power spectrum in the massless limit is zero,
and equals to  that of the massless  spin-$\frac12$  field.

The Correlation function is the Fourier transformation of  the power spectrum,
and will be presented in Sect. \ref{sec5}  and Appendix \ref{CCC} for the point-splitting scheme.

\section{Adiabatic regularization of stress tensor}\label{sec4}

In this section, we calculate the vacuum stress tensor of the spin-$\frac12$ field
using adiabatic regularization. We examine the ultraviolet structure of the
spectral energy density and pressure and show that the 2nd order subtraction
is sufficient to remove their UV divergences. In de Sitter space, integrating
the regularized spectra yields a maximally symmetric stress tensor with a
negative vacuum energy density for the massive field. We also examine the
massless limit and compare the 2nd and 4th order prescriptions, including
the dependence of the limiting result on the order of momentum integration
and the massless limit.

\begin{figure}[htb]
  \centering
       {%
          \includegraphics[width = .45\linewidth]{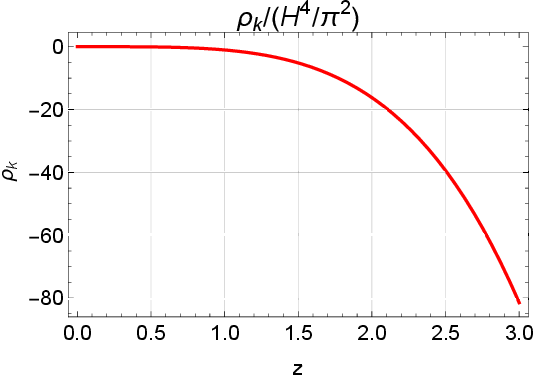}
  }
       {%
          \includegraphics[width = .45\linewidth]{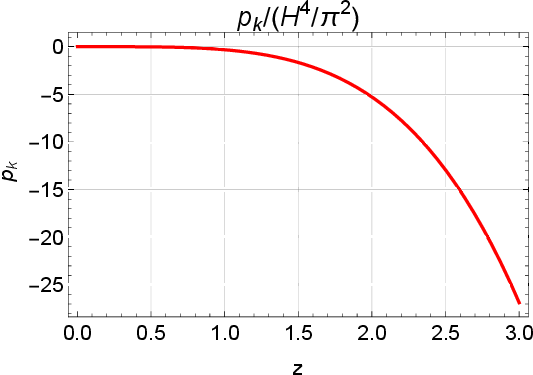}
      }
  \caption{
  (a) Unregularized spectral energy density $\rho_k$.
  (b) Unregularized spectral pressure $p_k$.
  }\label{unregenergyp}
  \end{figure}

The stress tensor of the field   $\psi$ in curved spacetimes
is defined by \cite{Birrell1982,Landete201314}
\bl
T_{\mu\nu}=\frac12i  [ \bar{\psi} \gamma_{(\mu}\nabla_{\nu)}\psi
- ( \nabla _{(\mu} \bar{\psi} )   \gamma_{\nu)} \psi ] .
\label{definationtmunu}
\el
The trace of  \eqref{definationtmunu}   is
\bl
T^{\mu}_{~\mu}  = g^{\mu\nu} T_{\mu\nu}= m\bar{\psi}\psi
, \label{traceofT}
\el
where  the Dirac equation \eqref{EOM} has been used for  the second equality.
Taking the vacuum expectation value of \eqref{definationtmunu}
gives the vacuum energy density and pressure as the following
\begin{subequations}
 \bl
\rho&=\langle0| T^{0}_{~~0}|0\rangle
=\int \frac{dk}{k}\rho_k,\label{srhoee}
\\
p&=-\frac13\langle0| T^{i}_{~~i}|0\rangle
=\int \frac{dk}{k}p_k,\label{peeee}
\el 
\end{subequations}
where
\begin{subequations}
\bl
\rho_k&=\frac{k^3}{2\pi^2a^3} i\big(h^{\Rmnum2}_k\dot{h}^{{\Rmnum2}*}_k
+h^{\Rmnum1}_k\dot{h}^{{\Rmnum1}*}_k
-\dot{h}^{\Rmnum2}_kh^{{\Rmnum2}*}_k-\dot{h}^{\Rmnum1}_kh^{{\Rmnum1}*}_k\big),
\label{rhotexpression}
\\
p_k&=\frac{k^4 }{2\pi^2 a^4}
(-\frac23)\big(h_k^{\Rmnum2}h_k^{\Rmnum1*}+h_k^{\Rmnum1}
h_k^{\Rmnum2*}\big) ,
\label{ptexpression}
\el
\end{subequations}
are the vacuum spectral energy density  and  spectral pressure, respectively.
By use of the equations \eqref{oneorderequ1} and \eqref{oneorderequ2},
the spectral energy density \eqref{rhotexpression} can be rewritten as
\bl
\rho_k&=-\frac{k^4}{\pi^2a^4} (h^{\Rmnum2}_kh^{\Rmnum1*}_k+h^{\Rmnum1}_kh^{\Rmnum2*}_k)
-m\frac{k^3}{\pi^2a^3}
(|h^{\Rmnum1}_k|^2-|h^{\Rmnum2}_k|^2).\label{EEE158main}
\el
From \eqref{EEE158main} it is seen that
\bl
\rho_k-3p_k=m \Delta^{2}_k,
\label{EEE159}
\el
where $p_k$ is given by  \eqref{ptexpression} and $\Delta^{2}_k$ is given by \eqref{400}.
Eq.\eqref{EEE159}  also follows
from the vacuum expectation value of eq.\eqref{traceofT}.
Using the modes  \eqref{exactlywavefunction1} and \eqref{exactlywavefunction2}
 into \eqref{rhotexpression} and \eqref{ptexpression} gives
\begin{subequations}
{\small
\bl
\rho_{k}
&=-i\frac{H^4}{\pi^2}z^4\frac{\sqrt{\pi z}}{4}e^{\pi\mu}
\Big(H^{(1)}_{-i\mu-\frac12}\frac{\partial }{\partial z}
(\frac{\sqrt{\pi z}}{2}H^{(2)}_{i\mu-\frac12})-H^{(2)}_{i\mu-\frac12}
\frac{\partial }{\partial z}(\frac{\sqrt{\pi z}}{2}H^{(1)}_{-i\mu-\frac12})\nn
\\
&~~~~~~~~~~~~~~~~
+H^{(1)}_{-i\mu+\frac12}\frac{\partial }{\partial z}
(\frac{\sqrt{\pi z}}{2}H^{(2)}_{i\mu+\frac12})-H^{(2)}_{i\mu+\frac12}
\frac{\partial }{\partial z}(\frac{\sqrt{\pi z}}{2}H^{(1)}_{-i\mu+\frac12})\Big),\label{Exactlyrho}
\\
p_{k}
&=\frac{H^4 }{2\pi^2 }z^4
\frac23i\frac{\pi z}{4}e^{\pi\mu}\Big(
      H^{(1)}_{-i\mu-\frac12}H^{(2)}_{i\mu+\frac12}-
H^{(2)}_{i\mu-\frac12}H^{(1)}_{-i\mu+\frac12}\Big).\label{Exactlyp}
\el}
\end{subequations}
Fig.\ref{unregenergyp} shows that
both $\rho_{k}$ and $p_{k}$ are negative, and UV divergent at  high $z$.
The negative sign of the unregularized vacuum spectral energy density
originates from the anticommutation relations of the spin-$\frac12$ field,
in contrast to the scalar and vector fields.

At high $z$, the spectral energy density and pressure are
\begin{subequations}
\bl
\lim_{z\rightarrow\infty}\rho_{k} &\simeq
 -\frac{H^4}{\pi^2}z^4 \Big(1+\frac{\mu ^2}{2z^2}
 -\frac{\mu ^2+\mu ^4}{8 z^4}+\frac{\mu ^2
 \left(4+5 \mu ^2+\mu ^4\right)}{16 z^6}
\nn
\\
& ~~~  -\frac{5 \mu ^2 \left(1+\mu ^2\right)
 \left(4+\mu ^2\right) \left(9+\mu ^2\right)}{128 z^8} \Big),
 \label{highkrhoexactly}
\\
\lim_{z\rightarrow\infty}p_{k} &\simeq
 \frac{H^4}{\pi^2}(-\frac13)z^4  \Big( 1-\frac{\mu ^2}{2 z^2}
 +\frac{3 \left(\mu ^2+\mu ^4\right)}{8 z^4}-\frac{5 \mu ^2
 \left(4+5 \mu ^2+\mu ^4\right)}{16 z^6}
 \nn \\
&~~~  +\frac{35 \mu ^2 \left(1+\mu ^2\right)
 \left(4+\mu ^2\right) \left(9+\mu^2\right)}{128 z^8} \Big),
 \label{highkpexactly}
\el
\end{subequations}
where the first three terms are, respectively,
quartic, quadratic, and logarithmically divergent.

At small $z$,  the spectral energy density and pressure are  infrared convergent,
\begin{subequations}
\bl
&\lim_{z\rightarrow0}\rho_{k}
 \simeq   - \frac{H^4}{2\pi^2}z^4  \big( \frac{2 \mu  \tanh(\pi  \mu )}{z}\big)<0,
 \label{lowzlimitrho}
\\
&\lim_{z\rightarrow0}p_{k}
 \simeq -\frac{H^4}{\pi^2}z^{4}
 Re   \Big( \frac{2^{-1+2 i \mu } z^{-2 i \mu }
 \Gamma(\frac{1}{2}+i \mu) }{3 \Gamma(\frac{1}{2}-i \mu) \cosh(\pi  \mu) }    \Big),
 \label{lowzlimitp}
\el
\end{subequations}
with  $\rho_k$  of \eqref{lowzlimitrho} being  negative,
and the sign of $p_k$  of \eqref{lowzlimitp} depending on the magnitude of $\mu$.

Now we shall remove the UV divergences in the unregularized $\rho_{k}$ and $p_{k}$.
In analogy to the power spectrum,
the   regularized spectral energy density and pressure are defined as the difference
\begin{subequations}
\bl
\rho_{k~reg}^{(n)}& \equiv \rho_{k}-\rho_{k~ad}^{(n)},~~ n=0,2,4,...
\\
p_{k~reg}^{(n)} &\equiv p_{k}-p_{k~ad}^{(n)}, ~~ n=0,2,4,...
  \label{nthregrhop}
\el
\end{subequations}
where   $\rho_{k~ad}^{(n)}$ and $p_{k~ad}^{(n)}$
are   the adiabatic spectral energy density and pressure,  listed in Appendix \ref{BBB}.
(See \eqref{B88}, \eqref{B11}, \eqref{B13},  \eqref{20000}, \eqref{20111}, \eqref{20222}.)
In  de Sitter space, they are given by the following
\begin{subequations}\bl
\rho^{(0)}_{k~ad}&=-\frac{H^4}{\pi^2}z^4\frac{1}{z}\bar{\omega},\label{zrhoad0th}
\\
p^{(0)}_{k~ad}
&=-\frac13\frac{H^4}{\pi^2}z^4
\frac{z}{\bar\omega},\label{zpad0th}
\\
\rho_{k~ad}^{(2)}
&=-\frac{
  H^4 }{\pi ^2}z^4\frac1z(\bar\omega-\frac{ \mu^2}{8 \bar\omega^3}+\frac{ \mu^4}{8 \bar\omega^5}),
\label{zrhoad2nd}
  \\
p_{k~ad}^{(2)}
&=-\frac13\frac{H^4z^4}{\pi ^2}\frac{z}{\bar\omega }\Big(1+\frac{3\mu ^2}{8 \bar\omega ^4 }
       -\frac{5\mu ^4}{8 \bar\omega ^6 }\Big)
,\label{zpad2nd}
  \\
\rho_{k~ad}^{(4)}
&=-\frac{ H^4 }{\pi ^2}z^4\frac1z(\bar\omega+\frac{ \mu^4}{8 \bar\omega^5}-\frac{ \mu^2}{8 \bar\omega^3}+\frac{\mu^2}{4 \bar\omega^5}-\frac{165 \mu^4}{128 \bar\omega^7}+\frac{119\mu^6}{64 \bar\omega^9}-\frac{105 \mu^8}{128 \bar\omega^{11}}),\label{zrhoad4th}
\\
p_{k~ad}^{(4)}
&=-\frac13\frac{H^4}{\pi ^2}z^4 \frac{z}{\bar\omega }\Big(1+\frac{3\mu ^2}{8 \bar\omega ^4 }
-\frac{5 \mu ^2(2+ \mu ^2)}{8 \bar\omega ^6 }+\frac{1155\mu ^4}{128 \bar\omega ^8 }
-\frac{1071 \mu ^6}{64 \bar\omega ^{10}}+\frac{1155\mu ^8}{128 \bar\omega ^{12} }\Big).
\label{zpad4th}
\el
\end{subequations}
The 0th order  \eqref{zrhoad0th} and \eqref{zpad0th} contain  a single divergent term,
  the 2nd  order \eqref{zrhoad2nd} and  \eqref{zpad2nd}
and the 4th order \eqref{zrhoad4th} and \eqref{zpad4th} contain two divergent terms.
To compare with
the unregularized \eqref{highkrhoexactly} and \eqref{highkpexactly} at high $z$,
we expand the adiabatic \eqref{zrhoad0th} - \eqref{zpad4th} at high $z$  as follows
\begin{subequations}
\bl
\lim_{z\rightarrow\infty}\rho_{k~ad}^{(0)}&\simeq-\frac{H^4}{\pi^2}z^4
\Big(1+\frac{\mu ^2}{2 z^2}-\frac{\mu ^4}{8 z^4}+\frac{\mu ^6}{16 z^6}-\frac{5 \mu ^8}{128 z^8}\Big),\label{highk0rho}
\\
\lim_{z\rightarrow\infty}p_{k~ad}^{(0)}&\simeq-\frac13\frac{H^4}{\pi^2}z^4
\Big(1-\frac{\mu ^2}{2 z^2}+\frac{3 \mu ^4}{8 z^4}-\frac{5 \mu ^6}{16 z^6}+\frac{35 \mu ^8}{128 z^8}\Big),\label{highk0p}
\\
\lim_{z\rightarrow\infty}\rho_{k~ad}^{(2)}&\simeq-\frac{H^4}{\pi^2}z^4
\Big(1+\frac{\mu ^2}{2 z^2}-\frac{\mu ^2 (1+\mu ^2)}{8 z^4}+\frac{\mu ^4 (5+\mu ^2)}{16 z^6}-\frac{5 \mu ^6 (14+\mu ^2)}{128 z^8}\Big),\label{highk2rho}
\\
\lim_{z\rightarrow\infty}p_{k~ad}^{(2)}&\simeq-\frac13\frac{H^4}{\pi^2}z^4
\Big(1-\frac{\mu ^2}{2 z^2}+\frac{3(\mu ^2+\mu ^4)}{8 z^4}-\frac{5 \mu ^4 (5+\mu ^2)}{16 z^6}+\frac{35 \mu ^6 (14+\mu ^2)}{128 z^8}\Big),\label{highk2p}
\\
\lim_{z\rightarrow\infty}\rho_{k~ad}^{(4)}&\simeq-\frac{H^4}{\pi^2}z^4
\Big(1+\frac{\mu ^2}{2 z^2}-\frac{\mu ^2 (1+\mu ^2)}{8 z^4}+\frac{\mu ^2 (4+5 \mu ^2+\mu ^4)}{16 z^6}-\frac{5 \mu ^4 (7+\mu ^2)^2}{128 z^8}\Big),\label{highk4rho}
\\
\lim_{z\rightarrow\infty}p_{k~ad}^{(4)}&\simeq-\frac13\frac{H^4}{\pi^2}z^4
\Big(1-\frac{\mu ^2}{2 z^2}+\frac{3 (\mu ^2+\mu ^4)}{8 z^4}-\frac{5 \mu ^2 (4+5 \mu ^2+\mu ^4)}{16 z^6}+\frac{35 \mu ^4 (7+\mu ^2)^2}{128 z^8}\Big).\label{highk4p}
\el
\end{subequations}
We calculate  the regularized spectral stress tensor for each order in the following.
The  0th order regularized spectral stress tensor  at high $z$ is given by
\begin{subequations}
\bl
\lim_{z\rightarrow\infty}\rho_{k~reg}^{(0)}&\simeq-\frac{H^4}{\pi^2}z^4\Big(-\frac{\mu ^2}{8 z^4}+\frac{\mu ^2 \left(4+5 \mu ^2\right)}{16 z^6}-\frac{5 \mu ^2 \left(36+49 \mu ^2+14 \mu ^4\right)}{128 z^8}\Big),\label{highkrhoreg0th}
\\
\lim_{z\rightarrow\infty}p_{k~reg}^{(0)}&\simeq-\frac13\frac{H^4}{\pi^2}z^4\Big(\frac{3 \mu ^2}{8 z^4}-\frac{5 \mu ^2 \left(4+5 \mu ^2\right)}{16 z^6}+\frac{35 \mu ^2 \left(36+49 \mu ^2+14 \mu ^4\right)}{128 z^8}\Big).\label{highkpreg0th}
\el
\end{subequations}
These expressions retain a logarithmic UV divergence, so the 0th order
subtraction is not sufficient.

The 2nd order regularized spectral stress tensor at high $z$ is given by
\begin{subequations}\bl
&\lim_{z\rightarrow\infty} \rho_{k~reg}^{(2)} \simeq
-\frac{H^4}{\pi^2}z^4\Big(\frac{\mu ^2}{4 z^6}-\frac{5 \mu ^2 \left(36+49 \mu ^2\right)}{128 z^8}\Big),\label{highkrhoreg2nd}
\\
&\lim_{z\rightarrow\infty}p_{k~reg}^{(2)}\simeq
-\frac13\frac{H^4}{\pi^2}z^4\Big(-\frac{5 \mu ^2}{4 z^6}+\frac{35 \mu ^2 \left(36+49 \mu ^2\right)}{128 z^8}\Big) ,\label{highkpreg2nd}
\el
\end{subequations}
being  UV convergent.
Furthermore, $\rho_{k~reg}^{(2)}$ remains negative at high $z$.
Thus, the 2nd order regularization  is sufficient
to remove all UV divergences in the spectral stress tensor,
and preserves the negative sign,
as for the power spectrum.

The  4th order regularized spectral stress tensor  at high $z$ is given by
\begin{subequations}
\bl
\lim_{z\rightarrow\infty}\rho_{k~reg}^{(4)}
  &\simeq  \frac{H^4}{\pi^2}z^4\Big(\frac{45 \mu ^2}{32 z^8}\Big)
  ,\label{highkrhoreg4th}
\\
\lim_{z\rightarrow\infty}p_{k~reg}^{(4)}
&\simeq-\frac13\frac{H^4}{\pi^2}z^4\Big(\frac{315 \mu ^2}{32 z^8}\Big)
 , \label{highkpreg4th}
\el
\end{subequations}
and $\rho_{k~reg}^{(4)}$ becomes positive.
Relative to the 2nd order result, the 4th order subtraction also removes
the UV convergent $z^{-2}$ terms. It therefore changes the finite spectral
remainder and its high momentum sign, although those additional terms
are not needed to cancel the UV divergences under the minimal subtraction
criterion adopted here.

\begin{figure}[htb]
\centering
    {%
        \includegraphics[width = .45\linewidth]{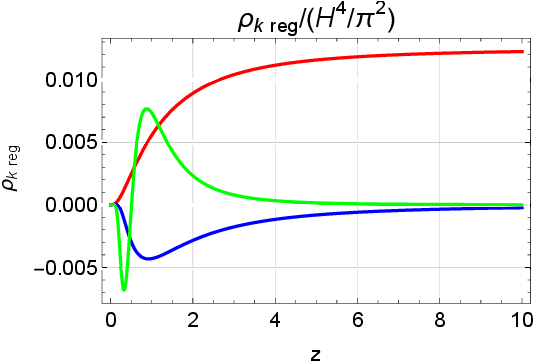}
}
    {%
        \includegraphics[width = .46\linewidth]{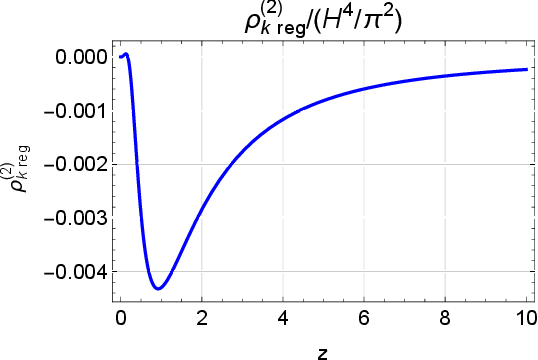}
    }
\caption{
(a) Regularized spectral energy density:
the 0th order $\rho_{k~reg}^{(0)}$ (red),
the 2nd order $\rho_{k~reg}^{(2)}$ (blue),
the 4th order $\rho_{k~reg}^{(4)}$ (green).
(b) The enlarged 2nd order  $\rho_{k~reg}^{(2)}$.
}
\label{rhok2}
\end{figure}

\begin{figure}[htb]
\centering
    {%
        \includegraphics[width = .45\linewidth]{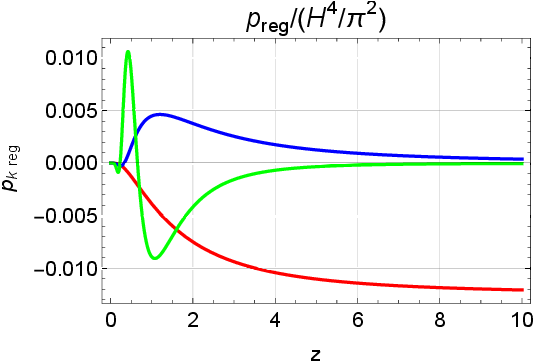}
}
    {%
        \includegraphics[width = .45\linewidth]{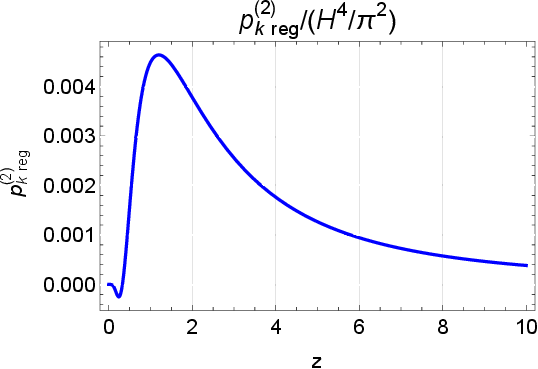}
    }
\caption{
(a) Regularized spectral pressure:
the 0th order $p_{k~reg}^{(0)}$ (red),
the 2nd order $p_{k~reg}^{(2)}$ (blue),
the 4th order $p_{k~reg}^{(4)}$ (green).
(b) The   enlarged 2nd order   $p_{k~reg}^{(2)}$.
}
\label{pk2}
\end{figure}

Fig.\ref{rhok2} (a) plots the  spectral energy density,
$\rho_{k~reg}^{(0)}$,  $\rho_{k~reg}^{(2)}$,  $\rho_{k~reg}^{(4)}$,
and Fig.\ref{pk2} (a) plots the spectral pressure, $p_{k~reg}^{(0)}$,  $p_{k~reg}^{(2)}$,  $p_{k~reg}^{(4)}$.
The enlarged {\small $\rho_{k~reg}^{(2)}$} and {\small $p_{k~reg}^{(2)}$}
are plotted in Fig.\ref{rhok2} (b) and Fig.\ref{pk2} (b).
It is seen that the 2nd order  $\rho_{k~reg}^{(2)}$ is negative except  at very small $z$,
and $p_{k~reg}^{(2)}$ is positive except at small $z$.
Like the power spectrum,
the infrared behavior of  $\rho_{k~reg}^{(2)}$ and $p_{k~reg}^{(2)}$
is due to the  inaccuracy of WKB modes at small $k$.

\begin{figure}[htb]
\centering
\includegraphics[width = .5\linewidth]{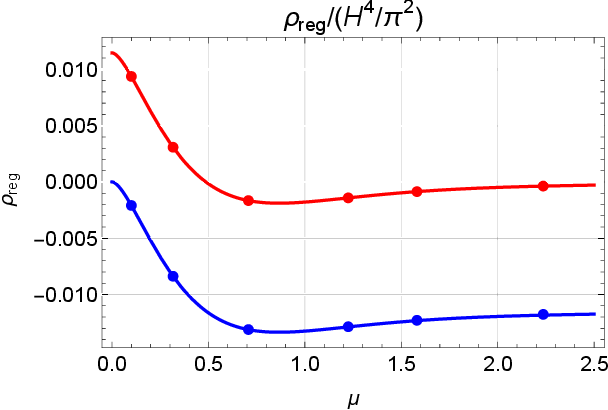}
\caption{
Blue line: the analytical 2nd order $\rho^{(2)}_{reg}$ of  \eqref{anrho} from the point-splitting.
Blue dots: the numerical  $\rho^{(2)}_{reg}$ of \eqref{rhoint2}  from the  adiabatic regularization.
Red line: the analytical 4th order $\rho^{(4)}_{reg}$ of  \eqref{rho4th}   from the point-splitting.
Red dots: the numerical  $\rho^{(4)}_{reg}$  from the adiabatic regularization.
}\label{fig6}
\end{figure}

At the spectral level, the regularized energy density and pressure do not
satisfy the vacuum equation of state mode by mode,
\bl
\rho_{k~reg}^{(2)} \neq -p_{k~reg}^{(2)} .
\el
Nevertheless,  integrating the spectra over $k$,
\bl
\rho_{reg}^{(2)} & \equiv \int_0^\infty\rho_{k~reg}^{(2)}\frac{dk}{k},
\label{rhoint2}
\\
p_{reg}^{(2)} & \equiv \int_0^\infty p_{k~reg}^{(2)}\frac{dk}{k} ,
\el
we find that the regularized  vacuum energy density and pressure
are  opposite to each other
\bl
\rho_{reg}^{(2)}  = - p_{reg}^{(2)} .
\label{rho2rl}
\el
For example, for    $\mu^2=0.1$,  the numerical integration  gives
\bl
\rho_{reg}^{(2)} & = -0.00837    \frac{H^4}{\pi^2} = - p_{reg}^{(2)}
 ,\label{rho2reg}
\el
and the numerical  $\rho_{reg}^{(2)}$   for other values of $\mu$
are plotted (in the blue dots) in Fig. \ref{fig6}.
It is remarkable that the outcome vacuum energy density and pressure  as in  \eqref{rho2rl}
are equal in magnitude  but with  opposite signs, $p_{reg}^{(2)}=-
\rho_{reg}^{(2)}$.
(For a simplified  model,  Ref \cite{Zeldovich1968} showed that
the finite, regularized  energy density and pressure given by
a generic scheme of regularization also satisfies the relation \eqref{rho2rl}.)
Thus,
the regularized vacuum stress tensor is proportional to the metric of background spacetime
\bl
\langle 0| T_{\mu\nu} |0\rangle_{reg}^{(2)}
 =\frac14g_{\mu\nu} \langle  0| T^{\beta}_{~~\beta} |0 \rangle_{reg}^{(2)}
    ,   \label{4th}
\el
and  possesses  the maximal symmetry in de Sitter space  \cite{weinberg1989}.
The massive spinor vacuum therefore has a maximally symmetric stress tensor,
but its regularized energy density is negative. It cannot by itself account
for the observed positive cosmological constant, although it may contribute
to the total vacuum energy together with other quantum fields. This differs
from the positive regularized vacuum energy densities obtained for the
massive scalar and vector fields considered in our previous work
\cite{yangzhang20201,pointsplittingyzw2022,XuanZhang2025}.
The primary result of the 2nd order regularization is thus a finite,
maximally symmetric vacuum stress tensor with negative energy density.

By numerical integration, we also find that the 4th order
$\rho_{reg}^{(4)}  = - p_{reg}^{(4)}$, like the 2nd order one,
and nevertheless that  $\rho_{reg}^{(4)}$ can be either positive or negative,
depending on the parameter $\mu$.
We plot the numerical  $\rho_{reg}^{(4)}$   (in the red dots) in Fig. \ref{fig6}.

The massless limit of the unregularized spectral stress tensor,  \eqref{Exactlyrho} and \eqref{Exactlyp},
is
\bl
\lim_{\mu\rightarrow0}\rho_{k}
&=\lim_{\mu\rightarrow0} 3 p_{k}
=-\frac{H^4}{\pi^2}z^4
   .\label{mlessrhoex}
\el
The massless limit of
the adiabatic spectral stress tensor, \eqref{zrhoad2nd}  \eqref{zpad2nd},
 is given by
\bl
&\rho^{(2)}_{k~ad} = 3 p^{(2)}_{k~ad} =
-\frac{H^4}{\pi^2}z^4,
\label{mlss2}
\el
equal to the unregularized \eqref{mlessrhoex},
so that the massless limit of  the regularized spectral stress tensor is vanishing,
\bl
\rho^{(2)}_{k~ reg} =p^{(2)}_{k~ reg}=0
   .  \label{trc2}
\el
It should be mentioned  that  \eqref{mlessrhoex}  - \eqref{trc2}
are actually also  valid for all the  orders ($n=0,2,4, ...$).
The massless limit of the regularized spectral stress tensor
is  equal to the regularized spectral stress tensor
of the massless spin-$\frac12$ field, both are zero.
Thus,  integrating  \eqref{trc2} over $k$, we get the zero trace
\bl
\langle 0| T^\mu_{~~\mu} |0\rangle^{(2)}_{reg}  =  \rho^{(2)}_{reg} - 3 p^{(2)}_{reg}=0.\label{edwdjwe}
\el
Thus, in the 2nd order subtraction considered here, the massless field
has a vanishing regularized stress tensor and trace, with no trace anomaly,
and gives no contribution to the cosmological constant. The continuity
of the integrated 2nd order result in the massless limit will also be
examined analytically in Sec.~\ref{sec5}.

For comparison, the 4th order regularization gives a nonzero limiting
stress tensor when the momentum integration is performed before taking
the massless limit. To examine this difference, we calculate the trace
difference between the 4th and 2nd order subtraction terms,
\bl
\lim_{m\rightarrow0}\int \big((\rho_{k~ad}^{(4)}-\rho_{k~ad}^{(2)})
   -3(p_{k~ad}^{(4)}-p_{k~ad}^{(2)})\big)\frac{dk}{k}
=\lim_{m\rightarrow0} m\int (\Delta_{k~ad}^{2(4)}-\Delta_{k~ad}^{2(2)})\frac{dk}{k}
,\label{dwedw}
\el
where the relation \eqref{EEE159} has been used.
From the adiabatic terms  \eqref{B3} and \eqref{B6},
the difference  reads
\bl
\Delta_{k~ad}^{2(4)}-\Delta_{k~ad}^{2(2)} =
 &-\frac{ k^3}{a^3\pi^2}\Big(
(\frac{ \dot{a}^4}{16 a^4 }+\frac{11 \dot{a}^2 \ddot{a}}{16 a^3 }
+\frac{7  \dot{a} \dddot{a}}{16 a^2 }+\frac{ \ddot{a}^2}{4 a^2 }
+\frac{ \ddddot{a}}{16 a })\frac{m}{\omega^5}
\nn
\\
& -(\frac{43  \dot{a}^4}{16 a^4 }+\frac{211 \dot{a}^2 \ddot{a}}{32a^3 }
+\frac{29  \ddot{a}^2}{32 a^2 }
+\frac{21\dot{a} \dddot{a}}{16 a^2}+\frac{ \ddddot{a}}{16 a})\frac{m^3}{\omega^7}\nn
\\
&    +(\frac{1659 \dot{a}^4}{128 a^4 }
+\frac{105  \dot{a}^2 \ddot{a}}{8 a^3 }+\frac{21  \ddot{a}^2}{32 a^2 }
+\frac{7\dot{a} \dddot{a}}{8 a^2 }
)\frac{m^5}{\omega^9}
\nn
\\
&  -(\frac{1239 m^7 \dot{a}^4}{64 a^4 \omega ^{11}}+\frac{231 m^7 \dot{a}^2 \ddot{a}}{32 a^3 \omega ^{11}})\frac{m^7}{\omega^{11}}
+\frac{1155 m^9 \dot{a}^4}{128 a^4 \omega ^{13}}
\Big).\label{dewdwe}
\el
Performing $k$-integration and using the formula
\[
\int_{0}^{\infty}\frac{x^2}{(1+x^2/b^2)^{\frac{n}2}}dx
=\frac{\sqrt{\pi}}{4} \frac{\Gamma(\frac{n}{2}-\frac32)}{\Gamma(\frac{n}{2})} b^3,
\]
one  gets
\bl
\int (\Delta_{k~ad}^{2(4)}-\Delta_{k~ad}^{2(2)})\frac{dk}{k}
& =  \frac{1}{240\pi^2 m}\Big(4\frac{\dot{a}^2}{a^2}\frac{\ddot{a}}{a}-3\frac{\ddot{a}^2}{a^2 }-9\frac{\dot{a}
      \dddot{a}}{a^2} -3\frac{\ddddot{a}}{a}\Big)
\\
&      = -   \frac{11 H^4 }{240 \pi^2 m }
 ,  \label{corr2}
\el
which is singular at zero mass.
Multiplying the above  by $m$ and taking the massless limit, one  gets
\bl
\lim_{m\rightarrow0}
 m\int (\Delta_{k~ad}^{2(4)}-\Delta_{k~ad}^{2(2)})\frac{dk}{k}
=  \frac{ 1}{240\pi^2}\Big(4\frac{\dot{a}^2}{a^2}\frac{\ddot{a}}{a}-3\frac{\ddot{a}^2}{a^2 }-9\frac{\dot{a}
      \dddot{a}}{a^2} -3\frac{\ddddot{a}}{a}\Big)
=       -   \frac{11 H^4 }{240 \pi^2  },\label{220}
\el
which corresponds to the trace anomaly discussed in Ref.~\cite{1Rio2014}. In this comparison, the finite term comes from the additional 4th order subtraction when the $k$-integration is performed before taking the massless limit. Taking the massless limit of the spectra first gives the vanishing result described above. Thus, momentum integration and the massless limit do not commute under the 4th order regularization.

The adiabatic calculation shows that the 2nd order subtraction is sufficient to remove the UV divergences of the vacuum spectral stress tensor. In de Sitter space, the regularized stress tensor of the massive spinor field is maximally symmetric and has negative energy density, while that of the massless field vanishes. The comparison with the 4th order regularization shows that the additional subtraction terms change the regularized spectra and affect the massless limit of the integrated quantities. In the next section, we derive Correlation functions and the corresponding subtraction terms. These expressions allow us to calculate the regularized stress tensor analytically using point-splitting regularization and check its agreement with the results from adiabatic regularization.

\section{Point-splitting regularization in coordinate space }\label{sec5}

The point-splitting regularization as a method works in  coordinate space
 \cite{DeWitt1975,Christensen1976, Christensen1978,pointsplittingyzw2022,yangzhang20201},
and can give the analytical, regularized correlation function and stress tensor,
whereas the adiabatic regularization in  $k$-space
can give the regularized power spectrum and spectral stress tensor.
The two methods are complementary.
We shall derive the analytic regularized correlation function and stress tensor,
and examine the property of the 4th order regularization in the massless limit.

The   unregularized  vacuum  correlation function  is defined by
\bl
  \langle 0| \bar\psi(x)\psi(x') |0\rangle
&  =  \frac{1}{a(t)^{\frac32}a(t')^{\frac32}|\vec{x}-\vec{x}'|}
\nn
\\
& \times \int_0^\infty \frac{k^3}{\pi^2} ( h_k^{\Rmnum2}(t) h_k^{\Rmnum2*}(t')
-h_k^{\Rmnum1}(t)h_k^{\Rmnum1*}(t'))
\frac{\sin k|\vec{x}-\vec{x}'|}{k} \frac{dk}{k}
       ,  \label{unrgr}
\el
and its coincidence limit (${x'\rightarrow x}$)
is the  UV divergent auto-correlation \eqref{selfcor}.
To remove the UV divergences,  one constructs  the adiabatic correlation function
\bl
 \langle 0| \bar\psi(x)\psi(x') |0 \rangle_{ad}^{(n)} ,
  ~~ n=0,2,4,...
  \label{2ndadiab}
\el
which is formed by using the   adiabatic modes $g^{I(n)}_k$ and $g^{II(n)}_k$
to replace the exact modes  $h^{I}_k$ and $h^{II}_k$ in \eqref{unrgr}.
Then one subtracts the adiabatic correlation from the unregularized correlation,
and takes the  coincidence limit,
\bl
  \langle  0| \bar\psi(x)\psi(x)  |0 \rangle_{reg}^{(2)}
\equiv    \lim_{x'\rightarrow x}\langle 0| \bar\psi(x)\psi(x') |0 \rangle
  - \lim_{x'\rightarrow x} \langle 0| \bar\psi(x)\psi(x') |0 \rangle_{ad}^{(2)}
  , \label{dehuiwh}
\el
where the 2nd order regularization is adopted.
 \eqref{dehuiwh} defines  the regularized auto-correlation
  in the point-splitting scheme,
and is analogous to the adiabatic regularization \eqref{reguidwelkn}
of the power spectrum in $k$-space.
From  the maximal symmetry \eqref{4th} and  the relation $T^{\beta}_{~~\beta}  = m\bar{\psi}\psi$,
the regularized vacuum stress tensor can be  expressed in terms of
the regularized  auto-correlation
\bl
\langle 0| T_{\mu\nu} |0\rangle_{reg}^{(2)}
     =   \frac14 g_{\mu\nu}       \,  m \langle 0| \bar\psi(x)\psi(x) |0 \rangle_{reg}^{(2)} .
\label{140tmunutrace}
\el
We first consider the simple case  of the massless field.
By the massless modes \eqref{exactlywavefunctionmasless1} and \eqref{exactlywavefunctionmasless2},
one has
\bl
h_k^{\Rmnum2}(t) h_k^{\Rmnum2*}(t') =h_k^{\Rmnum1}(t)h_k^{\Rmnum1*}(t'),
\el
so  the correlation function of the massless spin-$\frac12$ field   is zero
\bl
 \langle 0| \bar\psi(x)\psi(x') |0 \rangle= 0
 . \label{fjreio}
\el
Since the massless WKB modes \eqref{fhuier1} and \eqref{fhuier2}
are equal to  the exact modes,
 the adiabatic correlation function  is also zero
\bl
\langle 0| \bar\psi(x)\psi( x')  |0 \rangle^{(n)}_{ad} & =0,
   ~~~ n=0,2,4,...  , \label{fjreio2}
\el
which holds for  all adiabatic orders.
Thus,  the regularized  correlation  function of the massless field
 is zero
\bl
\lim_{x'\rightarrow x}
   \langle 0| \bar\psi(x)\psi( x') |0 \rangle^{(n)}_{reg}=0  ,
   ~~~ n=0,2,4,...
    \label{fhureihf}
\el
The result \eqref{fhureihf} is consistent with
the vanishing regularized power spectrum   \eqref{reglpower}
that has been derived from the adiabatic regularization.
The regularized  vacuum  stress tensor of the massless field is also zero
\bl
\langle 0| T_{\mu\nu} |0\rangle_{reg}^{(n)} =0 ,
         ~~~ n=0,2,4,...
\label{dyuw}
\el
This result from the point-splitting scheme
agrees with the results \eqref{trc2} and \eqref{edwdjwe} from the adiabatic scheme.

Next consider the massive case.
Inserting the  exact modes
 \eqref{exactlywavefunction1} and \eqref{exactlywavefunction2} into \eqref{unrgr}
and performing the integration,
we get the unregularized  correlation function (see Appendix \ref{CCC} for the details)
\bl
 \langle 0| \bar\psi(x)\psi(x') |0 \rangle
 &  =\frac{H^3}{\pi^2}\Gamma(\nu+\frac32)\Gamma(\frac32-\nu)\Big(
-\frac{i}4\nu \,_2F_1(\frac32-\nu,\frac32+\nu;2;1+\frac{\sigma_2}{2})
\nn
\\
&~~~~
+\frac{i}{16}(\frac94-\nu^2)
\,_2F_1(\frac{5}{2}-\nu,\frac{5}{2}+\nu;3;1+\frac{\sigma_2}{2})
\nn
\\
&~~~~
-\frac{i}{32}(\nu+\frac{5}{2})(\nu+\frac{3}{2})
\,_2F_1(\frac32-\nu,\frac72+\nu;3;1+\frac{\sigma_2}{2})
\nn
\\
&~~~~
-\frac{i}{32} (\frac{5}{2}-\nu)(\frac{3}{2}-\nu)
\,_2F_1(\frac72-\nu,\nu+\frac32;3;1+\frac{\sigma_2}{2})\Big)
  ,\label{unrcorr}
\el
where $\,_2F_1(a,b;c;d)$ is the hypergeometric function, $\nu\equiv-\frac12-i\mu$,
and
\bl
\sigma_2\equiv \frac12\frac{(a(t)-a(t'))^2}{a(t)a(t')}-\frac12a(t)a(t')H^2
|\vec{x}-\vec{x'}|^2
\label{sigma2}
\el
is one-half of the squared geodesic interval in de Sitter space. This correlation function is valid at all scales, unlike the short-distance expansions used to construct subtraction terms for the stress tensor \cite{Landete201314,Landete2013142}.

The 2nd order adiabatic correlation  function of the massive field can be derived as the following.
Use  the adiabatic modes $(g_k^{\Rmnum1(2)},~ g_k^{\Rmnum2(2)})$
to replace $(h_k^{\Rmnum1},~h_k^{\Rmnum2})$  in  \eqref{unrgr},
where the  integrand
\bl
 g_k^{\Rmnum2(2)}(t) g_k^{\Rmnum2(2)*}(t)
   -  g_k^{\Rmnum2(2)}(t) g_k^{\Rmnum2(2)*}(t)
\el
at the equal time ($t=t'$)
is the 2nd order adiabatic power spectrum $\Delta^{2(2)}_{k~ad}$ of \eqref{B3}.
Carrying out  the integration,
we obtain the 2nd order adiabatic correlation of the massive field
\bl
\langle 0| \bar\psi(x)\psi(x') |0 \rangle_{ad}^{(2)}
&=\frac{- H^3}{\pi^2}\frac{1}{\sqrt{-2\sigma_2}}
\Big(\mu^2K_1(\mu\sqrt{-2\sigma_2})
-\frac{1}{2}\mu\sqrt{-2\sigma_2}
K_0(\mu\sqrt{-2\sigma_2})\nn
\\
&~~~~
+\frac{3}{8}
(\mu\sqrt{-2\sigma_2})^2K_1(\mu\sqrt{-2\sigma_2})
-\frac{1}{24}(\mu\sqrt{-2\sigma_2})^3K_2(\mu\sqrt{-2\sigma_2})\Big)
 . \label{prooftrast2ad}
\el
In deriving \eqref{prooftrast2ad},
the following formula has been used \cite{SUM1980}
\bl
\int_0^\infty dz\frac{z \sin(zy)}{(z^2+\mu^2)^{n+\frac{1}{2}}}=\frac{-\sqrt{\pi}}{2^n\mu^n \Gamma(n+1/2)}
\frac{d}{dy} (y^n K_n(y\mu)) ,
\el
with $K_n(x)$ being the modified Bessel function and satisfying the relations
$\frac{d}{dx} K_0(x) = -K_1(x)$,
$\frac1x \frac{d}{dx}  (x K_1(x)) =  -  K_0(x)$,
$\frac1x \frac{d}{dx}  (x^2 K_2(x)) = - x K_1(x)$,
$\frac1x \frac{d}{dx}  (x^3 K_3(x)) = - x^2 K_2(x)$  \cite{Olver2010}.

We are more interested in the coincidence limit ($\sigma_2\rightarrow 0$).
The unregularized correlation  function \eqref{unrcorr} becomes
\bl
\lim_{\sigma_2\rightarrow0}\langle 0| \bar\psi(x)\psi(x') |0 \rangle
&\simeq\frac{H^3}{\pi^2}\Big(\frac{1}{2}\frac{\mu}{\sigma_2}
-\frac14\mu(1+\mu^2)(-1+2\gamma+\log(-\frac{\sigma_2 }{2})+\psi(2+i\mu)
+\psi(2-i\mu))\Big),\label{sTmumu}
\el
where $\psi$ on the rhs is the digamma function defined by $\psi(y) \equiv \frac{d}{dy} \ln \Gamma(y)$,
and a formula $\psi(1-i\mu)=\psi(2-i\mu)-\frac{1}{1-i \mu}$ has been used.
The 2nd order adiabatic correlation    \eqref{prooftrast2ad}  becomes
\bl
\lim_{\sigma_2\rightarrow0}\langle  0|  \bar\psi(x)\psi(x') |0 \rangle_{ad}^{(2)}
&\simeq \frac{H^3}{\pi^2} \Big(\frac{1}{2} \frac{\mu}{\sigma_2}
-\frac{ 13\mu }{24}-\frac{1}{4} \mu  (1+\mu^2)(-1+2\gamma+\log\mu^2
   +\log(-\frac{\sigma_2}{2}) ) \Big)
  . \label{u2nd}
\el
The difference between \eqref{sTmumu} and \eqref{u2nd}
gives  the 2nd  order regularized auto-correlation   function
\bl
 \langle  0|  \bar\psi(x)\psi(x)|0 \rangle_{reg}^{(2)}  & =
 \int   \Delta_{k~reg}^{2(2)}  \frac{dk}{k}
\nn \\
& = \frac{H^3}{\pi^2}\Big(
\frac{13}{24}\mu -\frac14\mu (1+\mu^2)(\psi(2+i\mu) +\psi(2-i\mu)-\log\mu^2)\Big)
  ,  \label{2ndautocor}
\el
where the UV divergences, $\frac{1}{\sigma_2}$ and $\log(-\frac{\sigma_2}{2})$,
have been subtracted off.
Multiplying  \eqref{2ndautocor} by   $\frac14 m g_{\mu\nu}$
yields  the 2nd  order regularized stress tensor
\bl
\langle  0|  T_{\mu\nu} |0\rangle^{(2)}_{reg}
&=\frac14g_{\mu\nu}
\frac{H^4}{\pi^2}\Big(
\frac{13}{24}\mu^2-\frac14\mu^2(1+\mu^2)(\psi(2+i\mu) +\psi(2-i\mu)-\log\mu^2)\Big)
   , \label{an}
\el
and the corresponding   regularized energy density and pressure are
\bl
\rho^{(2)}_{reg} = - p^{(2)}_{reg} =
   \frac14  \frac{H^4}{\pi^2}\Big( \frac{13}{24}\mu^2-\frac14\mu^2(1+\mu^2)(\psi(2+i\mu)
  +\psi(2-i\mu)-\log\mu^2)\Big) .
 \label{anrho}
\el
 \eqref{prooftrast2ad}  \eqref{2ndautocor}  \eqref{an}    are
our main result of the point-splitting scheme.
Given the expression,
we plot the analytical $\rho^{(2)}_{reg}$ (in the blue line)  vs the scaled mass $\mu$  in Fig.~\ref{fig6}.
 $\rho^{(2)}_{reg}$  is   negative, like the unregularized $\rho_k$.
For comparison, the numerical $\rho^{(2)}_{reg}$ (in the blue dots)
from the adiabatic regularization is also plotted in Fig.~\ref{fig6}.
The results from  the two schemes of regularization match consistently.

As  a consistency check,
the massless limit of  \eqref{unrcorr},  \eqref{prooftrast2ad}, \eqref{2ndautocor}, \eqref{an}
are  vanishing,
\begin{subequations}
 \bl
 \lim_{\mu =0 } \langle 0| \bar\psi(x)\psi(x') |0 \rangle & =  0,
 \\
 \lim_{\mu =0}
\langle 0| \bar\psi(x)\psi(x') |0 \rangle_{ad}^{(2)}  & =  0
, \label{adn}
\\
\lim_{\mu =0}   \langle 0| \bar\psi(x)\psi(x)|0 \rangle_{reg}^{(2)} &  =0
 ,  \label{2correg}
\\
 \lim_{\mu =0} \langle 0|  T_{\mu\nu} |0\rangle^{(2)}_{reg} & = 0,
\label{2stressreg}
\el 
\end{subequations}
agreeing  with \eqref{fjreio}, \eqref{fjreio2}, \eqref{fhureihf}, \eqref{dyuw}
of the massless field.
In particular, \eqref{2stressreg} shows that under the 2nd order regularization
the trace anomaly  never appears.
In computing  \eqref{2correg} \eqref{2stressreg},
we have used the following  formula for the di-gammar functions
\bl
\lim_{\mu\rightarrow0}(\psi(2+i\mu) +\psi(2-i\mu) )
                \simeq(2-2\gamma)   -  \mu^2 \psi^{(2)}(2) ,
\el
where the Euler number $\gamma \simeq 0.577$
and $\psi^{(2)}(2) \equiv d^2 \psi(z)/d z^2 |_{z=2} \simeq 0.404$.
From the above  it is seen  that, for  the 2nd order regularization,
the ordering of the massless limit and the $k$-integration can be  exchanged,
yielding the same  outcome,
\begin{subequations}\bl
\lim_{m\rightarrow0} \int   \Delta_{k~reg}^{2(2)}  \frac{dk}{k}
 & =  \int  \lim_{m\rightarrow0} \Delta_{k~reg}^{2(2)}  \frac{dk}{k} =0 ,
 \\
\lim_{m\rightarrow0} \int   \rho_{k~reg}^{(2)}  \frac{dk}{k}
 & =  \int  \lim_{m\rightarrow0} \rho_{k~reg}^{(2)}  \frac{dk}{k} =0 ,
\\
\lim_{m\rightarrow0} \int  p_{k~reg}^{(2)}  \frac{dk}{k}
 & =  \int  \lim_{m\rightarrow0} p_{k~reg}^{(2)}  \frac{dk}{k} =0 .
\el
\end{subequations}
The 4th order regularization is improper for the massive  spin-$\frac12$ field, as is known  in Sect 4.
Still we will reveal its difficulty via   the point-splitting scheme.
Similarly to the 2nd order case,
using the 4th order  adiabatic  modes $(g_k^{\Rmnum1(4)},~g_k^{\Rmnum2(4)})$
to replace $(h_k^{\Rmnum1},~h_k^{\Rmnum2})$  in  \eqref{unrgr},
and carrying out  the integration,
we get the  4th order adiabatic correlation     function
\bl
\langle 0| \bar\psi(x)\psi(x') |0 \rangle_{ad}^{(4)}
&=
\frac{-H^3}{\pi^2}\frac{1}{\sqrt{-2\sigma_2}}\Big[\nn
-\Big(\frac{17}{960} \mu ^3 (\sqrt{-2\sigma_2})^5+\frac{\mu ^3 }{24}(\sqrt{-2\sigma_2})^3+\frac{109 \mu  }{480}(\sqrt{-2\sigma_2})^3\nn
\\
&+\frac{\mu  }{2}(\sqrt{-2\sigma_2})\Big)K_{0}(\mu\sqrt{-2\sigma_2})+\Big(
\frac{\mu ^4 }{1152}(\sqrt{-2\sigma_2})^6+\frac{193 \mu ^2 }{1920}(\sqrt{-2\sigma_2})^4\nn
\\
&+\frac{7 \mu ^2 }{24}(\sqrt{-2\sigma_2})^2+\frac{11 }{240}(\sqrt{-2\sigma_2})^2
+\mu ^2\Big)K_{1}(\mu\sqrt{-2\sigma_2})  \Big]
, \label{4thdhewui}
\el
and its coincidence limit
\bl
\lim_{\sigma_2\rightarrow0}\langle  0|  \bar\psi(x)\psi(x')|0 \rangle_{ad}^{(4)}
&\simeq \frac{H^3}{\pi^2}\Big(\frac{1}{2}\frac{\mu}{\sigma_2}
-\frac{11}{240\mu}-\frac{13\mu}{24}
-\frac{1}{4} \mu  (1+\mu^2)(-1+2\gamma+\log\mu^2+\log(-\frac{\sigma_2}{2}))\Big)
  . \label{tT4th}
\el
The difference between \eqref{sTmumu} and \eqref{tT4th}
gives the 4th order regularized auto-correlation  
\bl
  \langle  0|  \bar\psi(x)\psi(x)|0 \rangle_{reg}^{(4)}
    & =  \int   \Delta_{k~reg}^{2(4)}  \frac{dk}{k}
\nn \\
&     =\frac{H^3}{\pi^2}\Big(\frac{11}{240 \mu}
    +\frac{13\mu}{24} -\frac14\mu (1+\mu^2)(\psi(2+i\mu) +\psi(2-i\mu)-\log\mu^2)\Big)
 \label{cor4th}
\\
 & =  \frac{H^3}{\pi^2}  \frac{11}{240 \mu}
  + \langle  0|  \bar\psi(x)\psi(x)|0 \rangle_{reg}^{(2)}
    ,  \label{4th2}
\el
which corresponds to   the result \eqref{corr2}.
Multiplying the above by   $\frac14 m g_{\mu\nu}$ yields the 4th order regularized stress tensor
\bl
\langle  0|  T_{\mu\nu} |0 \rangle^{(4)}_{reg}
&=\frac14 g_{\mu\nu}
   \frac{H^4}{\pi^2} \Big( \frac{11}{240}
  +\frac{13\mu^2}{24}
  -\frac14\mu^2(1+\mu^2)(\psi(2+i\mu) +\psi(2-i\mu)-\log\mu^2)\Big)
 \label{t4th}
\\
& =    g_{\mu\nu}    \frac{H^4}{\pi^2}  \frac{11}{960}
     + \langle  0|  T_{\mu\nu} |0 \rangle^{(2)}_{reg}
 . \label{a4th}
\el
Ref.~\cite{Landete2013142} derived  \eqref{t4th} by use of a regulator of integration,
without giving the full expressions  \eqref{unrcorr} \eqref{4thdhewui}.
The 4th order regularized energy density and pressure are
\bl
\rho^{(4)}_{reg} = - p^{(4)}_{reg} =
    \frac{H^4}{\pi^2} \frac{11}{960}  +\rho^{(2)}_{reg} .
\label{rho4th}
\el
Fig.~\ref{fig6} shows that  $\rho^{(4)}_{reg}$
is higher than  $\rho^{(2)}_{reg}$ by $\frac{H^4}{\pi^2} \frac{11}{960}$,
and becomes positive at small $\mu$.
This is due to the over-subtraction under the 4th order regularization.

Let us examine the difficulties associated with
 the massless limit of the 4th order regularization.
Firstly,  the massless limit of the  4th order regularized  auto-correlation  
\eqref{4th2}  is  singular
\bl
 \lim_{\mu =0 }
\langle 0| \bar\psi(x)\psi(x) |0 \rangle_{reg}^{(4)}
    = \frac{H^3}{\pi^2} \frac{11}{240 \mu} =  \infty  ,
\label{4thadiab}
\el
in contradiction   to
the zero correlation   function  \eqref{fhureihf} of the massless field.
Next, the massless limit of the 4th order regularized  stress tensor  \eqref{a4th} is
\bl
 \lim_{\mu =0 } \langle  0|  T_{\mu\nu} |0 \rangle^{(4)}_{reg}
 = g_{\mu\nu} \frac{H^4 }{ \pi^2} \frac{11}{960}
 , \label{4thadiabstr}
\el
in contradiction to the zero stress tensor \eqref{dyuw} of the  massless field, too.
So, for the 4th order regularization,
the ordering of the massless limit and the $k$-integration may not exchanged
\begin{subequations}\bl
\lim_{m\rightarrow0} \int   \Delta_{k~reg}^{2(4)}  \frac{dk}{k}
 & \ne  \int  \lim_{m\rightarrow0} \Delta_{k~reg}^{2(4)}  \frac{dk}{k} =0 ,
 \\
\lim_{m\rightarrow0} \int   \rho_{k~reg}^{(4)}  \frac{dk}{k}
 & \ne \int  \lim_{m\rightarrow0} \rho_{k~reg}^{(4)}  \frac{dk}{k} =0,
\\
\lim_{m\rightarrow0} \int  p_{k~reg}^{(4)}  \frac{dk}{k}
 & \ne \int  \lim_{m\rightarrow0} p_{k~reg}^{(4)}  \frac{dk}{k} =0 ,
\el
\end{subequations}
unlike the 2nd order case.
The trace anomaly will appear only in the 4th order regularization
with the $k$-integration preceding  the massless limit \cite{Landete201314,Landete2013142,1Rio2014},
but will disappear when the massless limit is taken first.
These differences suggest limitations of the 4th order regularization in recovering the massless field results from the massive spin-$\frac12$ field.

\section{Conclusion and Discussion}\label{sec6}

We have studied the vacuum regularization of spinor fields in de Sitter space using both adiabatic and point-splitting regularization. The adiabatic expansion of the spinor mode functions was first revisited. The conditions commonly used in the literature do not uniquely fix several arbitrary functions appearing in the WKB expansion. Instead of setting these functions to zero, they were kept explicitly in the calculation. Their contributions cancel in the adiabatic power spectrum and spectral stress tensor up to the 4th order. The physical quantities are therefore independent of these arbitrary functions, which also explains why setting them to zero in previous calculations gives the same results.

The ultraviolet structure of the power spectrum and spectral stress tensor was then examined. The 2nd order adiabatic regularization is sufficient to remove all ultraviolet divergences for the massive spinor field in a general spatially flat Robertson-Walker spacetime. The 0th order is not sufficient, while higher-order subtraction is not required for removing the divergences. In de Sitter space, the 2nd order regularized stress tensor of the massive field is maximally symmetric, although its spectral form is not uniform in momentum space. The regularized energy density remains negative, in agreement with the sign of the unregularized vacuum energy density. For the massless field, the 2nd order regularized power spectrum and stress tensor vanish, together with the trace.

Analytic correlation   functions for both massive and massless spinor fields in de Sitter space were also derived, together with their 2nd and 4th order adiabatic correlation functions. These expressions are valid at all scales and allow the point-splitting regularization to be carried out analytically in coordinate space. The resulting regularized correlation function and stress tensor agree with those obtained from numerical integration of the adiabatically regularized spectra. The agreement between the two methods shows that the momentum-space and coordinate-space calculations lead to the same regularized stress tensor.

The behavior in the massless limit further distinguishes the different regularization orders. The exact and 2nd order adiabatic correlation functions of the massive field reduce smoothly to those of the massless field as the mass approaches zero. The same is true for the 2nd order regularized stress tensor, which reduces to the vanishing stress tensor of the massless field. By contrast, the 4th order regularized quantities do not preserve this smooth limit. In particular, the 4th order subtraction can change the sign and low-momentum behavior of the spectral energy density, and the resulting stress tensor does not reduce continuously to the massless result. The trace anomaly appears under this 4th order prescription when the momentum integration is performed before taking the massless limit, while it is absent in the 2nd order regularization and for a field that is massless from the beginning.

These results also have implications for the possible vacuum origin of the cosmological constant. The regularized stress tensor of the massless spinor field vanishes and therefore gives no contribution to the cosmological constant. For the massive field, the regularized stress tensor is maximally symmetric, but its vacuum energy density is always negative. A massive spinor field alone therefore cannot account for the observed positive cosmological constant. It may nevertheless contribute to the total vacuum energy together with other quantum fields whose regularized vacuum energy densities are positive, such as massive scalar and vector fields \cite{yangzhang20201,pointsplittingyzw2022,XuanZhang2025}. The cosmological constant may thus depend on the combined vacuum contributions of different quantum fields.

\

\

\textbf{Acknowledgements}

\

Xuan Ye is in part by NSFC Grant No. 12433002. Yang Zhang is supported by NSFC Grant No. 12261131497.

\

\appendix
\numberwithin{equation}{section}

\section{WKB modes }\label{AAA}

We shall derive the WKB modes $g_k^{\Rmnum1}$ of \eqref{45} and $g_k^{\Rmnum2}$ of \eqref{46}
 up to the 4th adiabatic orders.
Our treatments on the arbitrary functions  of $\Omega$, $F$, and $G$
are different from those in Refs. \cite{Landete201314, Landete2013142, 1Rio2014}.

Replacing the exact functions $h^{I}_k$ and $h^{II}_k$ with the WKB functions
 $g^{I}_k$ and $g^{II}_k$ in
\eqref{normalcon}, \eqref{oneorderequ1} and \eqref{oneorderequ2} yields
\bl
&g_k^{\Rmnum1}(t) =i\frac{a}{k}(\partial_0-im) g_k^{\Rmnum2}(t),\label{48111}
\\
&g_k^{\Rmnum2}(t)
= i\frac{a}{k}(\partial_0 +i m )g_k^{\Rmnum1}(t),\label{48222}
\\
&|g_k^{\Rmnum1}(t)|^2+|g_k^{\Rmnum2}(t)|^2=1,\label{48333}
\el
Plugging \eqref{45} and \eqref{46} into \eqref{48111},\eqref{48222}, and \eqref{48333},
one gets the equations of $\Omega$, $F$ and $G$ as the following
\bl
&\Omega G+ i \dot{G} +i\frac{G}{2}\frac{d\omega}{dt}\left(\frac{1}{\omega-m}-\frac{1}{\omega }\right)+m G=(\omega+m)F,\label{49}
\\
&\Omega F+ i \dot{F} +i\frac{F}{2}\frac{d\omega}{dt}\left(\frac{1}{\omega+m}-\frac{1}{\omega }\right)-m F=(\omega-m)G,\label{50}
\\
&(\omega+m)FF^*+(\omega-m)GG^*=2\omega,\label{51}
\el
agreeing with (15) in \cite{1Rio2014}.
Decompose $F$ and $G$ into the real and imaginary parts as the following
\bl
F&\equiv RF+i IF =\sum_{n} (RF^{(n)}+iIF^{(n)}), \label{deihwiu}
\\
G&\equiv RG+i IG =\sum_n (RG^{(n)}+iIG^{(n)}) . \label{dnweoid}
\el
Then the relations \eqref{relFGn1} and \eqref{relFGn2} lead  to the following
\bl
RF^{(n)}(t; -m)&=RG^{(n)}(t; m),\label{RIGFn11}
\\
IF^{(n)}(t; -m)&=IG^{(n)}(t; m),\label{RIGFn22}
\\
\omega^{(n)}(t; -m)&=\omega^{(n)}(t; m) .\label{RIGFn}
\el
Plugging \eqref{deihwiu} and \eqref{dnweoid} into \eqref{49}, \eqref{50}, and \eqref{51}
yields
\bl
&\Omega RG- \dot{IG} -\frac{IG}{2}\frac{d\omega}{dt}\left(\frac{1}{\omega-m}-\frac{1}{\omega }\right)+m RG=(\omega+m)RF,\label{53}
\\
&\Omega RF- \dot{IF} -\frac{IF}{2}\frac{d\omega}{dt}\left(\frac{1}{\omega+m}-\frac{1}{\omega }\right)-m RF=(\omega-m)RG,\label{54}
\\
&\Omega IG+ \dot{RG} +\frac{RG}{2}\frac{d\omega}{dt}\left(\frac{1}{\omega-m}-\frac{1}{\omega }\right)+m IG=(\omega+m)IF,\label{55}
\\
&\Omega IF+ \dot{RF} +\frac{RF}{2}\frac{d\omega}{dt}\left(\frac{1}{\omega+m}-\frac{1}{\omega }\right)-m IF=(\omega-m)IG,\label{56}
\\
&(\omega+m)(RF^2+IF^2)+(\omega-m)(RG^2+IG^2)=2\omega.\label{E13}
\el
In the following  we shall solve the set of equations \eqref{53} $\sim$ \eqref{E13}
order by order.
Substituting \eqref{47}, \eqref{deihwiu}, and \eqref{dnweoid}
into \eqref{53} $\sim$ \eqref{E13},
we get  the following, for the respective order,

{\bf 0th order:}

\bl
(\omega+m)&=(\omega+m),
\\
(\omega-m)&=(\omega-m),
\\
0&=0,
\\
0&=0,
\\
(\omega+m)+(\omega-m)&=2\omega.\label{0thequtions}
\el
which are the identities.

{\bf 1st order:}

\bl
&\omega^{(1)} +\omega RG^{(1)}+m RG^{(1)}=(\omega+m)RF^{(1)},\label{61}
\\
&\omega^{(1)} +\omega RF^{(1)}-m RF^{(1)}=(\omega-m)RG^{(1)},\label{62}
\\
&\omega IG^{(1)} -\frac{1}{2}\frac{\dot a}{a}\frac{m (m+\omega )}{\omega ^2}+m IG^{(1)}=(\omega+m)IF^{(1)},\label{63}
\\
&\omega IF^{(1)}-\frac{1}{2}\frac{\dot a}{a}\frac{m (m-\omega )}{\omega ^2}-m IF^{(1)}=(\omega-m)IG^{(1)},\label{64}
\\
& (RF^{(1)}+RG^{(1)})+\frac{m}{\omega}(RF^{(1)}-RG^{(1)})=0,\label{65}
\el
where $IG^{(0)}=RG^{(0)}=1$ (see \eqref{47})
 and {\small$d\omega/ dt=\frac{\dot{a}}{a}\frac{ 1}{\omega  }(m^2-\omega^2)
$} have been used.
 \eqref{63} and \eqref{64} give
\bl
IG^{(1)}-IF^{(1)}=\frac12\frac{\dot{a}}{a}\frac{m}{\omega^2}.\label{66}
\el
Simplifying \eqref{61} and \eqref{62} yields
\bl
RG^{(1)}-RF^{(1)}&=-\frac{\omega^{(1)}}{m+\omega},\label{EE23plusone}
\\
RG^{(1)}-RF^{(1)}&=-\frac{\omega^{(1)}}{m-\omega},\label{EE23}
\el
which imply
\bl
\omega^{(1)}&=0,\label{A22plusone}
\\
RG^{(1)}-RF^{(1)}&=0.\label{A22}
\el
Combining \eqref{65} with \eqref{A22} leads to
\bl
RG^{(1)}=RF^{(1)}=0.\label{E26}
\el
so  $F^{(1)}$ and $G^{(1)}$  are imaginary and can be written as  \cite{Landete201314}
\bl
F^{(1)}&=i(-A\frac{\dot{a}}{a}\frac{m}{\omega^2}+K\frac{\dot{a}}{a}),
\label{A33}
\\
G^{(1)}&=i(B\frac{\dot{a}}{a}\frac{m}{\omega^2}+K\frac{\dot{a}}{a}),
\label{A34}
\el
where $(A,B,K)$ are some real functions depending on $m$ and $\omega$
with the appropriate dimensions.
\eqref{66} leads to a constraint $A+B=\frac12$.
In fact,  $A=B=\frac14$, as we shall see later from the 2nd adiabatic order.
Thus, \eqref{A33} and \eqref{A34} become
\bl
F^{(1)}=-\frac14i\frac{\dot{a}}{a}\frac{m}{\omega^2}+iK\frac{\dot{a}}{a},~~~~
G^{(1)}=\frac14i\frac{\dot{a}}{a}\frac{m}{\omega^2}+iK\frac{\dot{a}}{a}.\label{EEE35}
\el
By the relation \eqref{RIGFn22}, one has  $K(t; m)=K(t; -m)$,
so $K $ is  even in $m$ and can be nonzero in general.
Our calculation differs from Ref. \cite{Landete2013142} which assumed $IF^{(1)}(m)=-IG^{(1)}(m)$.

{\bf 2nd order}

\bl
&\omega^{(2)}- \dot{IG}^{(1)} -\frac{IG^{(1)}}{2}\frac{d\omega}{dt}\left(\frac{1}{\omega-m}-\frac{1}{\omega }\right)=(\omega+m)(RF^{(2)}-RG^{(2)}),\label{80}
\\
&\omega^{(2)} - \dot{IF}^{(1)} -\frac{IF^{(1)}}{2}\frac{d\omega}{dt}\left(\frac{1}{\omega+m}-\frac{1}{\omega }\right)=(\omega-m)(RG^{(2)}-RF^{(2)}),\label{81}
\\
&\omega IG^{(2)}+m IG^{(2)}=(\omega+m)IF^{(2)},\label{82}
\\
&\omega IF^{(2)}-m IF^{(2)}=(\omega-m)IG^{(2)},\label{83}
\\
&(\omega+m)(2RF^{(2)}+IF^{(1)2})+(\omega-m)(2RG^{(2)}+IG^{(1)2})=0,\label{84}
\el
where $\omega^{(1)}=RG^{(1)}=RF^{(1)}=IG^{(0)}=IF^{(0)}=0$
have been used.
$\eqref{82}$ and $\eqref{83}$  yield the  equation
\bl
IG^{(2)}=IF^{(2)}.\label{EE54}
\el
Ref.~\cite{Landete201314}  set $K = IF^{(2)} = IG^{(2)} = 0$ based on
an assumption $IF^{(n)}(t; m) = - IG^{(n)}(t; m)$.
But, as we see, only the relation $IF^{(2)}(t; m) = IG^{(2)}(t; -m)$ will follow  from \eqref{relFGn1},
and $IG^{(2)}$ and $IF^{(2)}$ can be nonzero  in general.
Eqs. \eqref{80}$\sim$\eqref{84} reduce to the following
inhomogeneous linear equations,
\bl
&\omega^{(2)}-(\omega+m)(RF^{(2)}-RG^{(2)}) = \dot{IG}^{(1)}+\frac{IG^{(1)}}{2}\frac{\dot{a}}{a}\frac{ 1}{\omega  }\left(m^2-\omega^2\right)\left(\frac{1}{\omega-m}-\frac{1}{\omega }\right),\label{86}
\\
&\omega^{(2)} -(\omega-m)(RG^{(2)}-RF^{(2)})=\dot{IF}^{(1)} +\frac{IF^{(1)}}{2}\frac{\dot{a}}{a}\frac{ 1}{\omega  }\left(m^2-\omega^2\right)\left(\frac{1}{\omega+m}-\frac{1}{\omega }\right),\label{87}
\\
&2(\omega+m)RF^{(2)}+2(\omega-m)RG^{(2)}
=-(\omega+m)IF^{(1)2}-(\omega-m)IG^{(1)2},\label{88}
\el
and the solutions  are
\bl
\omega^{(2)}&=K\left(\frac{ \ddot{a}}{a}-\frac{ \dot{a}^2}{a^2}\right)+\dot{K}\frac{\dot{a} }{a}+(2A-\frac12)\left(\frac{m^3 \dot{a}^2}{a^2 \omega ^4}-\frac{m \dot{a}^2}{2 a^2 \omega ^2}-\frac{m \ddot{a}}{2 a \omega ^2}\right)+\left(\frac{5 m^4 \dot{a}^2}{8 a^2 \omega^5}-\frac{3 m^2 \dot{a}^2}{8 a^2 \omega ^3}-\frac{m^2 \ddot{a}}{4 a \omega^3}\right),\label{EE51}
\\
RF^{(2)}&=-K\left(K -2A \frac{ m }{ \omega ^2}\right)\frac{\dot{a}^2}{2a^2}-A^2\frac{m^2 \dot{a}^2}{2 a^2 \omega ^4}+\frac{ m^2 R}{48\omega ^4}-\frac{5 m^4 \dot{a}^2}{16 a^2 \omega ^6}-\frac{m R }{48\omega ^3}+\frac{5 m^3 \dot{a}^2}{16 a^2 \omega ^5},\label{EE52}
\\
RG^{(2)}&=-K\left(K -(2A-1) \frac{ m }{ \omega ^2}\right)\frac{\dot{a}^2}{2a^2}-\left(A^2-A+\frac{1}{4}\right)\frac{m^2 \dot{a}^2}{2 a^2 \omega ^4}+\frac{ m^2 R}{48\omega ^4}-\frac{5 m^4 \dot{a}^2}{16 a^2 \omega ^6}+\frac{m R }{48\omega ^3}-\frac{5 m^3 \dot{a}^2}{16 a^2 \omega ^5},\label{EE53}
\el
where $B=\frac12-A$ has been used,
and  $R=6(\frac{\ddot{a}}{a}+\frac{\dot{a}^2}{a^2})$ is the Ricci scalar.
Imposing the constraint \eqref{RIGFn} on \eqref{EE51}
and the constraint \eqref{RIGFn11} on \eqref{EE52} and \eqref{EE53},
we find
\bl
A=\frac14. \label{EE57}
\el
Then,  \eqref{EE51}, \eqref{EE52}, and \eqref{EE53} become
\bl
\omega^{(2)}&=K\left(\frac{ \ddot{a}}{a}-\frac{ \dot{a}^2}{a^2}\right)+\omega\frac{dK}{d\omega}(\frac{m^2}{\omega^2}-1)\frac{\dot{a}^2}{a^2}+\frac{5 m^4 \dot{a}^2}{8 a^2 \omega^5}-\frac{3 m^2 \dot{a}^2}{8 a^2 \omega ^3}-\frac{m^2 \ddot{a}}{4 a \omega^3},\label{89}
\\
RF^{(2)}&=-K\left(K -\frac12\frac{ m }{ \omega ^2}\right)\frac{\dot{a}^2}{2a^2}+\frac{ m^2 R}{48\omega ^4}-\frac{5 m^4 \dot{a}^2}{16 a^2 \omega ^6}-\frac{m^2 \dot{a}^2}{32 a^2 \omega ^4}-\frac{m R }{48\omega ^3}+\frac{5 m^3 \dot{a}^2}{16 a^2 \omega ^5},\label{90}
\\
RG^{(2)}&=-K\left(K +\frac12\frac{ m }{ \omega ^2}\right)\frac{\dot{a}^2}{2a^2}+\frac{ m^2 R}{48\omega ^4}-\frac{5 m^4 \dot{a}^2}{16 a^2 \omega ^6}-\frac{m^2 \dot{a}^2}{32 a^2 \omega ^4}+\frac{m R }{48\omega ^3}-\frac{5 m^3 \dot{a}^2}{16 a^2 \omega ^5},\label{91}
\el
where $\dot{K}=\frac{d K}{d\omega}\frac{d\omega}{dt}$  has been used.

{\bf 3rd order}

\bl
&(\omega+m)( RG^{(3)}-RF^{(3)})+\omega^{(3)}= \dot{IG}^{(2)} +\frac{IG^{(2)}}{2}\frac{d\omega}{dt}\left(\frac{1}{\omega-m}-\frac{1}{\omega }\right),\label{EE75}
\\
&(\omega-m)( RF^{(3)}-RG^{(3)})+\omega^{(3)}= \dot{IF}^{(2)} +\frac{IF^{(2)}}{2}\frac{d\omega}{dt}\left(\frac{1}{\omega+m}-\frac{1}{\omega }\right),\label{EE76}
\el
\bl
&(\omega +m)(IG^{(3)}-IF^{(3)})=-\omega^{(2)}IG^{(1)}- \dot{RG}^{(2)} -\frac{RG^{(2)}}{2}\frac{d\omega}{dt}\left(\frac{1}{\omega-m}-\frac{1}{\omega }\right),\label{EE77}
\\
&(\omega-m) (IF^{(3)}-IG^{(3)})=-\omega^{(2)}IF^{(1)}- \dot{RF}^{(2)} -\frac{RF^{(2)}}{2}\frac{d\omega}{dt}\left(\frac{1}{\omega+m}-\frac{1}{\omega }\right),\label{EE78}
\el
\bl
&(\omega+m)(RF^{(3)}+IF^{(1)}IF^{(2)})
+(\omega-m)(RG^{(3)}+IG^{(1)}IG^{(2)})=0,\label{EE79}
\el
where the relations $\omega^{(1)}=RG^{(1)}=RF^{(1)}=IG^{(0)}=IF^{(0)}=0$ have
been used. Solving  \eqref{EE77} and \eqref{EE78} yields
\bl
IF^{(3)}-IG^{(3)}&=\frac{65 m^5 \dot{a}^3}{32 a^3 \omega ^8}-\frac{97 m^3 \dot{a}^3}{64 a^3 \omega ^6}+\frac{m \dot{a}^3}{8 a^3 \omega ^4}-\frac{19 m^3 \dot{a} \ddot{a}}{16 a^2 \omega ^6}+\frac{m \dot{a} \ddot{a}}{2 a^2 \omega ^4}+\frac{m \dddot{a} }{8 a \omega ^4}\nn
\\
&~~~~~~~~~~~~~~~~~~~~~~~~~~+K(\frac{5  m^3 \dot{a}^3}{8 a^3 \omega ^5}-\frac{ m \dot{a}^3}{4 a^3 \omega ^3}+\frac{K m \dot{a}^3}{4 a^3 \omega ^2}-\frac{ m \dot{a} \ddot{a}}{4 a^2 \omega ^3}).\label{EE80}
\el
By $IF^{(3)}(t; -m)=IG^{(3)}(t; m)$, we can write $IF^{(3)}$ and $IG^{(3)}$   as
\bl
IF^{(3)}&=\frac{65 m^5 \dot{a}^3}{64 a^3 \omega ^8}-\frac{97 m^3 \dot{a}^3}{128 a^3 \omega ^6}+\frac{m \dot{a}^3}{16 a^3 \omega ^4}-\frac{19 m^3 \dot{a} \ddot{a}}{32 a^2 \omega ^6}+\frac{m  \dot{a} \ddot{a}}{4 a^2 \omega ^4}+\frac{m \dddot{a}}{16 a \omega ^4}\nn
\\
&~~~~~~~~~~~~~~+K(\frac{5  m^3  \dot{a}^3}{16 a^3 \omega ^5}-\frac{ m \dot{a}^3}{8 a^3 \omega ^3}+\frac{K m \dot{a}^3}{8 a^3 \omega ^2}-\frac{ m \dot{a} \ddot{a}}{8 a^2 \omega ^3})+L\frac{\dddot{a}}{a}+M\frac{\ddot{a}\dot{a}}{a^2}+N\frac{\dot{a}^3}{a^3},\nn
\\
IG^{(3)}&=-\frac{65 m^5 \dot{a}^3}{64 a^3 \omega ^8}+\frac{97 m^3 \dot{a}^3}{128 a^3 \omega ^6}-\frac{m \dot{a}^3}{16 a^3 \omega ^4}+\frac{19 m^3 \dot{a} \ddot{a}}{32 a^2 \omega ^6}-\frac{m \dot{a} \ddot{a}}{4 a^2 \omega ^4}-\frac{m \dddot{a}}{16 a \omega ^4}\nn
\\
&~~~~~~~~~~~~~~~-K(\frac{5  m^3 \dot{a}^3}{16 a^3 \omega ^5}-\frac{ m \dot{a}^3}{8 a^3 \omega ^3}+\frac{K m \dot{a}^3}{8 a^3 \omega ^2}-\frac{ m \dot{a}\ddot{a}}{8 a^2 \omega ^3})
+L\frac{\dddot{a}}{a}+M\frac{\dot{a}\ddot{a}}{a^2}+N\frac{\dot{a}^3}{a^3},\label{EEE81}
\el
where $L$, $M$, and $N$  are arbitrary functions even in $m$,
\bl
L(t; -m)&=L(t; m),
\\
M(t; -m)&=M(t; m),
\\
N(t; -m)&=N(t; m),
\el
and can be nonzero in general.
These  arbitrary functions were set to zero in Refs. \cite{Landete2013142,1Rio2014}.

From \eqref{EE75}, \eqref{EE76}, and \eqref{EE79},
we have the following
\bl
\omega^{(3)}&=\dot{IF}^{(2)},\label{A53}
\\
RF^{(3)} &=-IF^{(2)} \left( K -\frac{ m }{4 \omega^2}\right)\frac{\dot{a}}{a }
=-IF^{(2)}IF^{(1)},\label{116}
\\
RG^{(3)} &=-IG^{(2)} \left( K  +\frac{ m }{4 \omega^2}\right)\frac{\dot{a}}{a }
=-IG^{(2)}IG^{(1)} ,\label{1116}
\el
where \eqref{EEE35} has been used.
In general,  $RF^{(3)}$ and $RG^{(3)}$  can be  nonzero.

{\bf  4th order}

\bl
&(\omega RG^{(4)}+\omega^{(2)}RG^{(2)}+\omega^{(4)})- \dot{IG}^{(3)} -\frac{IG^{(3)}}{2}\frac{d\omega}{dt}\left(\frac{1}{\omega-m}-\frac{1}{\omega }\right)+m RG^{(4)}=(\omega+m)RF^{(4)},\label{129}
\\
&(\omega RF^{(4)}+\omega^{(2)}RF^{(2)}+\omega^{(4)})- \dot{IF}^{(3)} -\frac{IF^{(3)}}{2}\frac{d\omega}{dt}\left(\frac{1}{\omega+m}-\frac{1}{\omega }\right)-m RF^{(4)}=(\omega-m)RG^{(4)},\label{130}
\\
&(\omega^{(2)}IG^{(2)}+\omega^{(3)}IG^{(1)})
+ \dot{RG}^{(3)} +\frac{RG^{(3)}}{2}\frac{d\omega}{dt}\left(\frac{1}{\omega-m}-\frac{1}{\omega }\right)=(\omega+m)(IF^{(4)}-IG^{(4)}),\label{E97}
\\
&(\omega^{(2)}IF^{(2)}+\omega^{(3)}IF^{(1)})+ \dot{RF}^{(3)} +\frac{RF^{(3)}}{2}\frac{d\omega}{dt}\left(\frac{1}{\omega+m}-\frac{1}{\omega }\right)=(\omega-m)(IG^{(4)}-IF^{(4)}),\label{E98}
\\
&(\omega+m)(2RF^{(4)}+2IF^{(1)}IF^{(3)}+RF^{(2)}RF^{(2)}+IF^{(2)}IF^{(2)}),\nn
\\
&~~~~~~~~~~~~~~~~~~~~~~~~~
+(\omega-m)(2RG^{(4)}+2IG^{(1)}IG^{(3)}+RG^{(2)}RG^{(2)}+IG^{(2)}IG^{(2)})=0,\label{124}
\el
where we have used
$IF^{(0)}=IG^{(0)}=0,~RF^{(0)}=RG^{(0)}=1,
~\omega^{(1)}=RF^{(1)}=RG^{(1)}=0$.
$\eqref{E97}$ and $\eqref{E98}$ lead to  the following  equation
\bl
&IF^{(4)}-IG^{(4)}=IF^{(2)}\Big(\frac{5 m^3 \dot{a}^2}{8 a^2 \omega^5}
-\frac{m \dot{a}^2}{4 a^2 \omega^3}+K\frac{m  \dot{a}^2}{2a^2 \omega^2 }-\frac{m \ddot{a}}{4 a \omega^3}\Big),\label{E103}
\el
where $IF^{(4)}$ and $IG^{(4)}$ remain undetermined.
Solving the equations  \eqref{129}, \eqref{130}, and \eqref{124},  one gets
\bl
\omega^{(4)}&=-\frac{1105 m^8 \dot{a}^4}{128 a^4 \omega ^{11}}+\frac{337 m^6 \dot{a}^4}{32 a^4 \omega ^9}-\frac{377 m^4 \dot{a}^4}{128 a^4 \omega ^7}+\frac{3 m^2 \dot{a}^4}{32 a^4 \omega ^5}+\frac{221 m^6 \dot{a}^2 \ddot{a}}{32 a^3 \omega ^9}-\frac{389 m^4 \dot{a}^2 \ddot{a}}{64 a^3 \omega ^7}+\frac{13 m^2 \dot{a}^2 \ddot{a}}{16 a^3 \omega ^5}-\frac{19 m^4 \ddot{a}^2}{32 a^2 \omega ^7}\nn
\\
&+\frac{m^2 \ddot{a}^2}{4 a^2 \omega ^5}-\frac{7 m^4 \dot{a} a^{(3)}}{8 a^2 \omega ^7}+\frac{15 m^2 \dot{a} \dddot{a}}{32 a^2 \omega ^5}+\frac{m^2 \ddddot{a}}{16 a \omega ^5}\nn
\\
&+K\frac{23 m^4  \dot{a}^2 \ddot{a}}{16 a^3 \omega ^6}-K\frac{17 m^2  \dot{a}^2 \ddot{a}}{32 a^3 \omega ^4}-K\frac{3 m^2 \dot{a}^4}{32 a^4 \omega ^4}+K^3\frac{ \dot{a}^2 \ddot{a}}{2 a^3}-M\frac{2 \dot{a}^2 \ddot{a}}{a^3}+N\frac{3  \dot{a}^2 \ddot{a}}{a^3}-K\frac{m^2  \ddot{a}^2}{8 a^2 \omega ^4}-L\frac{ \dot{a}\dddot{a}}{a^2}+M\frac{\dot{a}\dddot{a}}{a^2}\nn
\\
&-K\frac{m^2  \dot{a}\dddot{a}}{8 a^2 \omega ^4}+M\frac{\ddot{a}^2}{a^2}+L\frac{ \ddddot{a}}{a}-K^3\frac{ \dot{a}^4}{2 a^4}+K\frac{21 m^4  \dot{a}^4}{16 a^4 \omega ^6}-N\frac{3  \dot{a}^4}{a^4}-K\frac{15 m^6  \dot{a}^4}{8 a^4 \omega ^8}+K\dot{K}\frac{ \dot{a}^3 }{2 a^3}+\dot{M}\frac{\dot{a} \ddot{a}}{a^2}+\dot{N}\frac{\dot{a}^3 }{a^3}\nn
\\
&+\dot{L}\frac{ \dddot{a}}{a}+\dot{K}\frac{5 m^4 \dot{a}^3 }{16 a^3 \omega ^6}
-\dot{K}\frac{3 m^2 \dot{a}^3 }{32 a^3 \omega ^4}
-\dot{K}\frac{m^2 \dot{a} \ddot{a}}{8 a^2 \omega ^4},\label{A67}
\el

\allowbreak{
\bl
&RF^{(4)}=+\frac{2285 m^8 \dot{a}^4}{512 a^4 \omega ^{12}}-\frac{565 m^7 \dot{a}^4}{128 a^4 \omega ^{11}}-\frac{1263 m^6 \dot{a}^4}{256 a^4 \omega ^{10}}+\frac{2611 m^5 \dot{a}^4}{512 a^4 \omega ^9}
+\frac{2371 m^4 \dot{a}^4}{2048 a^4 \omega ^8}-\frac{333 m^3 \dot{a}^4}{256 a^4 \omega ^7}
-\frac{3 m^2 \dot{a}^4}{128 a^4 \omega ^6}
+\frac{m \dot{a}^4}{32 a^4 \omega ^5}
\nn
\\
&-\frac{457 m^6 \dot{a}^2 \ddot{a}}{128 a^3 \omega ^{10}}+\frac{113 m^5 \dot{a}^2 \ddot{a}}{32 a^3 \omega ^9}+\frac{725 m^4 \dot{a}^2 \ddot{a}}{256 a^3 \omega ^8}-\frac{749 m^3 \dot{a}^2 \ddot{a}}{256 a^3 \omega ^7}-\frac{19 m^2 \dot{a}^2 \ddot{a}}{64 a^3 \omega ^6}+\frac{11 m \dot{a}^2 \ddot{a}}{32 a^3 \omega ^5}+\frac{41 m^4 \ddot{a}^2}{128 a^2 \omega ^8}\nn
\\
&-\frac{5 m^3 \ddot{a}^2}{16 a^2 \omega ^7}-\frac{17 m^2 \ddot{a}^2}{128 a^2 \omega ^6}+\frac{m \ddot{a}^2}{8 a^2 \omega ^5}+\frac{7 m^4 \dot{a} \dddot{a}}{16 a^2 \omega ^8}-\frac{7 m^3 \dot{a} \dddot{a}}{16 a^2 \omega ^7}-\frac{13 m^2 \dot{a} \dddot{a}}{64 a^2 \omega ^6}+\frac{7 m \dot{a} \dddot{a}}{32 a^2 \omega ^5}-\frac{m^2 \ddddot{a}}{32 a \omega ^6}+\frac{m \ddddot{a}}{32 a \omega ^5}\nn
\\
&-\frac{1}{2} IF^{(2)2}-K^4\frac{ \dot{a}^4}{8 a^4}-KN \frac{\dot{a}^4}{a^4}-K \frac{15 m^5 \dot{a}^4}{16 a^4 \omega ^8}+K \frac{47 m^3 \dot{a}^4}{64 a^4 \omega ^6}- K^2\frac{5 m^4 \dot{a}^4}{32 a^4 \omega ^6}
-K^2 \frac{5 m^3 \dot{a}^4}{32 a^4 \omega ^5}
\nn
\\
& -K\frac{m\dot{a}^4}{16 a^4 \omega ^4}
+K^2\frac{3 m^2\dot{a}^4}{64 a^4 \omega ^4}
+K^2 \frac{m \dot{a}^4}{16 a^4 \omega ^3}
+N\frac{m\dot{a}^4}{4 a^4 \omega ^2}
-K M\frac{ \dot{a}^2 \ddot{a}}{a^3}
+K\frac{9 m^3\dot{a}^2 \ddot{a}}{16 a^3 \omega ^6}
-K\frac{m\dot{a}^2 \ddot{a}}{4 a^3 \omega ^4}
\nn
\\
&  +K^2\frac{m^2\dot{a}^2 \ddot{a}}{16 a^3 \omega ^4}
+K^2\frac{m\dot{a}^2 \ddot{a}}{16 a^3 \omega ^3}
+M\frac{m\dot{a}^2 \ddot{a}}{4 a^3 \omega ^2}
-K L \frac{\dot{a} \dddot{a}}{a^2}
-K \frac{m \dot{a} \dddot{a}}{16 a^2 \omega ^4}
+L\frac{m\dot{a} \dddot{a}}{4 a^2 \omega ^2},
\label{A65}
\el
}
\bl
RG^{(4)}= & +\frac{2285 m^8 \dot{a}^4}{512 a^4 \omega ^{12}}+\frac{565 m^7 \dot{a}^4}{128 a^4 \omega ^{11}}-\frac{1263 m^6 \dot{a}^4}{256 a^4 \omega ^{10}}-\frac{2611 m^5 \dot{a}^4}{512 a^4 \omega ^9}+\frac{2371 m^4 \dot{a}^4}{2048 a^4 \omega ^8}+\frac{333 m^3 \dot{a}^4}{256 a^4 \omega ^7}
\nn
\\
& -\frac{3 m^2 \dot{a}^4}{128 a^4 \omega ^6}
-\frac{m \dot{a}^4}{32 a^4 \omega ^5}
-\frac{457 m^6 \dot{a}^2 \ddot{a}}{128 a^3 \omega ^{10}}
-\frac{113 m^5 \dot{a}^2 \ddot{a}}{32 a^3 \omega ^9}
+\frac{725 m^4 \dot{a}^2 \ddot{a}}{256 a^3 \omega ^8}
+\frac{749 m^3 \dot{a}^2 \ddot{a}}{256 a^3 \omega ^7}
\nn \\
& -\frac{19 m^2 \dot{a}^2 \ddot{a}}{64 a^3 \omega ^6}
 -\frac{11 m \dot{a}^2 \ddot{a}}{32 a^3 \omega ^5}
+\frac{41 m^4 \ddot{a}^2}{128 a^2 \omega ^8}
+\frac{5 m^3 \ddot{a}^2}{16 a^2 \omega ^7}
-\frac{17 m^2 \ddot{a}^2}{128 a^2 \omega ^6}
-\frac{m \ddot{a}^2}{8 a^2 \omega ^5}
\nn
\\
&
+\frac{7 m^4 \dot{a} \dddot{a}}{16 a^2 \omega ^8}
+\frac{7 m^3 \dot{a} \dddot{a}}{16 a^2 \omega ^7}
-\frac{13 m^2 \dot{a} \dddot{a}}{64 a^2 \omega ^6}-\frac{7 m \dot{a} \dddot{a}}{32 a^2 \omega ^5}-\frac{m^2 \ddddot{a}}{32 a \omega ^6}-\frac{m \ddddot{a}}{32 a \omega ^5}
\nn
\\
&-\frac{1}{2} IF^{(2)2}-K^4\frac{ \dot{a}^4}{8 a^4}-KN \frac{\dot{a}^4}{a^4}
+K \frac{15 m^5 \dot{a}^4}{16 a^4 \omega ^8}
-K \frac{47 m^3 \dot{a}^4}{64 a^4 \omega ^6}- K^2\frac{5 m^4 \dot{a}^4}{32 a^4 \omega ^6}
+K^2 \frac{5 m^3 \dot{a}^4}{32 a^4 \omega ^5}
\nn
\\
& +K\frac{m\dot{a}^4}{16 a^4 \omega ^4}
+K^2\frac{3 m^2\dot{a}^4}{64 a^4 \omega ^4}-K^2 \frac{m \dot{a}^4}{16 a^4 \omega ^3}-N\frac{m\dot{a}^4}{4 a^4 \omega ^2}-K M\frac{ \dot{a}^2 \ddot{a}}{a^3}-K\frac{9 m^3\dot{a}^2 \ddot{a}}{16 a^3 \omega ^6}
+K\frac{m\dot{a}^2 \ddot{a}}{4 a^3 \omega ^4}
\nn
\\
& +K^2\frac{m^2\dot{a}^2 \ddot{a}}{16 a^3 \omega ^4}
-K^2\frac{m\dot{a}^2 \ddot{a}}{16 a^3 \omega ^3}
-M\frac{m\dot{a}^2 \ddot{a}}{4 a^3 \omega ^2}-K L \frac{\dot{a} \dddot{a}}{a^2}
+K \frac{m \dot{a} \dddot{a}}{16 a^2 \omega ^4}-L\frac{m\dot{a} \dddot{a}}{4 a^2 \omega ^2},
\label{A66}
\el
which contain arbitrary   functions $K$, $L$, $M$, $N$, $IF^{(2)}$.

We have shown that, some arbitrary functions appear in the WKB modes at each order,
and can not be completely determined by the conditions \eqref{RIGFn11}, \eqref{RIGFn22}, and \eqref{RIGFn}.
Nevertheless,  as we shall show in Appendix \ref{BBB},
these arbitrary functions will cancel out in
the power spectrum and the spectral stress tensor.

\section{Adiabatic spectra}\label{BBB}

Using  the WKB modes given in Appendix \ref{AAA},
we shall calculate the adiabatic power spectrum and spectral stress tensor,
and show that  the arbitrary functions cancel out in the results.

{\bf Adiabatic power spectrum}

The formula of  power spectrum is \eqref{400}.
The adiabatic  power spectrum is given by
using the WKB modes $g^{I }_k$ and $g^{II}_k$
of \eqref{45} and \eqref{46} to replace  $h^{I }_k$ and $h^{II }_k$.
Keeping terms up to each order, we get the following, respectively,
\bl
\Delta^{2(0)}_{k~ad}&=-\frac{ k^3}{a^3\pi^2}\big(\frac{\omega+m}{2\omega}-\frac{\omega-m}{2\omega})=-\frac{ k^3}{a^3\pi^2}\frac{m}{\omega},\label{B1}
\\
\Delta^{2(1)}_{k~ad} &= \Delta^{2(0)}_{k~ad} ,
\\
\Delta^{2(2)}_{k~ad}&=-\frac{ H^3}{\pi^2}z^3\big(\frac{\omega+m}{2\omega}(IF^{(1)2}+2RF^{(2)})
-\frac{\omega-m}{2\omega}(IG^{(1)2}+2RG^{(2)})\big)+\Delta^{2(0)}_{k~ad}\nn
\\
&=-\frac{ k^3}{a^3\pi^2}\big(\frac{m}{\omega}-\frac{5 m^5 \dot{a}^2}{8 a^2 \omega ^7}
+\frac{7 m^3 \dot{a}^2}{8 a^2 \omega ^5}-\frac{m \dot{a}^2}{4 a^2 \omega ^3}
+\frac{m^3 \ddot{a}}{4 a \omega ^5}-\frac{m \ddot{a}}{4 a \omega ^3}\big)
 ,\label{B3}
\\
\Delta^{2(3)}_{k~ad} &= \Delta^{2(2)}_{k~ad} ,
\el
being   independent of the arbitrary functions $K$, $IF^{(2)}$, $IG^{(2)}$,
\bl
\Delta^{2(4)}_{k~ad}
= & -\frac{ k^3}{a^3\pi^2} \Big(\frac{\omega+m}{2\omega}2(
RF^{(4)}+IF^{(1)} IF^{(3)}+\frac12(IF^{(2)2}+RF^{(2)2}))\nn
\\
&~~~~~~~~~~~  -\frac{\omega-m}{2\omega}2(
RG^{(4)}+IG^{(1)} IG^{(3)}+\frac12(IG^{(2)2}+RG^{(2)2}))\Big)+\Delta^{2(2)}_{k~ad}
\nn
\\
= & -\frac{ k^3}{a^3\pi^2}\Big(
\frac{m}{\omega}-\frac{5 m^5 \dot{a}^2}{8 a^2 \omega ^7}+\frac{7 m^3 \dot{a}^2}{8 a^2 \omega ^5}
-\frac{m \dot{a}^2}{4 a^2 \omega ^3}+\frac{m^3 \ddot{a}}{4 a \omega ^5}-\frac{m \ddot{a}}{4 a \omega ^3}
\nn
\\
&+\frac{1155 m^9 \dot{a}^4}{128 a^4 \omega ^{13}}-\frac{1239 m^7 \dot{a}^4}{64 a^4 \omega ^{11}}+
\frac{1659 m^5 \dot{a}^4}{128 a^4 \omega ^9}-\frac{43 m^3 \dot{a}^4}{16 a^4 \omega ^7}
+\frac{m \dot{a}^4}{16 a^4 \omega ^5}-\frac{231 m^7 \dot{a}^2 \ddot{a}}{32 a^3 \omega ^{11}}
\nn \\
& +\frac{105 m^5 \dot{a}^2 \ddot{a}}{8 a^3 \omega ^9}
 -\frac{211 m^3 \dot{a}^2 \ddot{a}}{32 a^3 \omega ^7}
 +\frac{11 m \dot{a}^2 \ddot{a}}{16 a^3 \omega ^5}+\frac{21 m^5 \ddot{a}^2}{32 a^2 \omega ^9}
 -\frac{29 m^3 \ddot{a}^2}{32 a^2 \omega ^7}+\frac{m \ddot{a}^2}{4 a^2 \omega ^5}
\nn \\
& +\frac{7 m^5 \dot{a} \dddot{a}}{8 a^2 \omega ^9}
- \frac{21 m^3 \dot{a} \dddot{a}}{16 a^2 \omega ^7}
+\frac{7 m \dot{a} \dddot{a}}{16 a^2 \omega ^5}
-\frac{m^3 \ddddot{a}}{16 a \omega ^7}
+\frac{m \ddddot{a}}{16 a \omega^5}\Big)
,\label{B6}
\el
being independent of  the arbitrary functions $K, L, M, N$,   $IG^{(2)},IF^{(2)}, IG^{(4)},IF^{(4)}$.

{\bf Adiabatic spectral stress tensor   }

We now compute the adiabatic spectral stress tensor up to the 4th  order,
and  show that it is independent of the arbitrary functions appearing in the WKB modes.

The formula of the spectral pressure is \eqref{ptexpression}.
The adiabatic  spectral pressure is given by
using the WKB modes $g^{I }_k$ and $g^{II )}_k$ to replace  $h^{I }_k$ and $h^{II }_k$.
To each  order, we get the following,  respectively
\bl
 p^{(0)}_{k~ad} & =\frac{k^4 }{2\pi^2 a^4} (-\frac23)\frac{\sqrt{\omega^2-m^2}}{\omega},\label{B88}
 \\
  p^{(1)}_{k~ad} & = p^{(0)}_{k~ad} ,
\el

\bl
p_{k~ad}^{(2)}
&=\frac{k^4 }{2\pi^2 a^4}
(-\frac23)\frac{\sqrt{\omega^2-m^2}}{\omega}(RF^{(2)}+RG^{(2)}+
IG^{(1)}IF^{(1)}+RG^{(1)}RF^{(1)})
+p_k^{(0)}\nn
\\
&=\frac{k^4 }{2\pi^2 a^4}
(-\frac23)\frac{\sqrt{\omega^2-m^2}}{\omega}\Big(1-\frac{5 m^4 \dot{a}^2}{8 a^2 \omega ^6}
+\frac{ m^2 \dot{a}^2}{8 a^2 \omega ^4}+\frac{m^2 \ddot{a}}{4 a \omega ^4}
\Big),\label{B11}
\\
p^{(3)}_{k~ad} & = p^{(2)}_{k~ad} ,
\el
being  independent of  the arbitrary functions $K$, $IF^{(2)}$, $IG^{(2)}$,

\bl
p_{k~ad}^{(4)} = & \frac{k^4 }{2\pi^2 a^4}
(-\frac23)\frac{\sqrt{\omega^2-m^2}}{\omega} \big(RF^{(4)}+RG^{(4)}+IF^{(2)2}+
IF^{(3)} IG^{(1)}+ IF^{(1)} IG^{(3)}+ RF^{(2)} RG^{(2)}  \big)
 +p_k^{(3)}\nn
\\
= & \frac{k^4 }{2\pi^2 a^4}
(-\frac23)\frac{\sqrt{\omega^2-m^2}}{\omega}\Big(1-\frac{5 m^4 \dot{a}^2}{8 a^2 \omega ^6}+\frac{ m^2 \dot{a}^2}{8 a^2 \omega ^4}+\frac{m^2 \ddot{a}}{4 a \omega ^4}
\frac{1155 m^8 \dot{a}^4}{128 a^4 \omega ^{12}}-\frac{609 m^6 \dot{a}^4}{64 a^4 \omega ^{10}}
\nn \\
&  +\frac{259 m^4 \dot{a}^4}{128 a^4 \omega ^8}-\frac{m^2 \dot{a}^4}{32 a^4 \omega ^6}
 -\frac{231 m^6 \dot{a}^2 \ddot{a}}{32 a^3 \omega ^{10}}+\frac{175 m^4 \dot{a}^2 \ddot{a}}{32 a^3 \omega ^8}
-\frac{m^2 \dot{a}^2 \ddot{a}}{2 a^3 \omega ^6}+\frac{21 m^4 \ddot{a}^2}{32 a^2 \omega ^8}
\nn \\
& -\frac{9 m^2 \ddot{a}^2}{32 a^2 \omega ^6}
+\frac{7 m^4 \dot{a} \dddot{a}}{8 a^2 \omega ^8}
-\frac{3 m^2 \dot{a} \dddot{a}}{8 a^2 \omega ^6}
-\frac{m^2 \ddddot{a}}{16 a \omega ^6}\Big),\label{B13}
\el
being  independent of the arbitrary functions
$K$, $L$, $M$, $N$, $IF^{(2)}$, $IG^{(2)}$, $IF^{(4)}$, $IG^{(4)}$.

The adiabatic spectral energy density can be written as
$\rho_{k~ad}^{(n)}=m\Delta^{2(n)}_{k~ad}+3p_{k~ad}^{(n)}$ according to the relation \eqref{EEE159}.
So we get
\bl
&\rho^{(0)}_{k~ad}=-\frac{k^3}{\pi^2a^3}\omega,\label{20000}
\\
&\rho_{k~ad}^{(2)}
=-\frac{k^3}{\pi^2a^3}\Big(\omega +(\frac{m^4 }{8 \omega ^5}-\frac{m^2}{8  \omega^3})
\frac{\dot{a}^2}{a^2}\Big),\label{20111}
\\
&\rho_{k~ad}^{(4)}
=-\frac{k^3}{\pi^2a^3}\Big(\omega +(\frac{m^4 }{8 \omega ^5}-\frac{m^2}{8  \omega ^3})
\frac{\dot{a}^2}{a^2}-(\frac{105 m^8 }{128  \omega ^{11}}-\frac{91 m^6 }{64 \omega ^9}
+\frac{81 m^4 }{128 \omega ^7}-\frac{m^2 }{32 \omega ^5})\frac{\dot{a}^4}{a^4}
\nn
\\
&~~~ +(\frac{7 m^6 }{16  \omega ^9}-\frac{5 m^4 }{8 a^3 \omega ^7}+\frac{3 m^2 }{16  \omega ^5})
\frac{\dot{a}^2 \ddot{a}}{a^3}+(\frac{m^4 }{32 \omega ^7}-\frac{m^2 }{32  \omega ^5})
\frac{\ddot{a}^2}{a^2 }-(\frac{m^4 }{16\omega ^7}-\frac{m^2 }{16  \omega ^5})\frac{\dot{a} \dddot{a}}{a^2}
\Big),\label{20222}
\el
which are independent of the arbitrary functions, too.

Thus, to  the each   order,  the arbitrary functions cancel out in
the adiabatic  power spectrum  and in the adiabatic  spectral stress tensor.
Therefore, in practice, these functions can be set to zero,
$K=L=M=N= IF^{(2)}=IG^{(2)}=IF^{(4)}=IG^{(4)}=0$,
as in Refs. \cite{Landete201314,Landete2013142,1Rio2014}.

 We have also verified that the  adiabatic spectral stress tensor is conserved, to each order,
\bl
\dot{\rho}_{k~ad}^{(n)}+3\frac{\dot{a}}{a}(\rho_{k~ad}^{(n)}
+p_{k~ad}^{(n)})=0, ~~~~~  n=0,2,4 .
\el
So the regularized spectral stress tensor is also conserved  to each order.

\section{The correlation  function in de Sitter space}\label{CCC}

In this appendix, we derive the analytic expression   \eqref{unrcorr}
of  the unregularized correlation function in de Sitter space.
The integration involved
is similar to that for the scalar field  \cite{BunchDavies1978,yangzhang20201}.
We first consider the equal-time case $t=t'$ for convenience,
and extend the result to the general case of $t\neq t'$
by using  the maximal symmetry of de Sitter space.
Plugging the modes $h_k^{\Rmnum1}(z)$ of \eqref{exactlywavefunction1}
and  $h_k^{\Rmnum2}(z)$ of \eqref{exactlywavefunction2}   into
 the  correlation   function \eqref{unrgr} yields
\bl
\langle 0| \bar\psi(x)\psi(x') |0 \rangle
& =\frac{-i}{4\pi H a^{4}|\vec{x}-\vec{x}'|}\int_0^\infty
  k^2\Big(    H^{(1)}_{-i\mu-\frac12}(z)   H^{(2)}_{-i\mu+\frac12}(z)
            +H^{(1)}_{-i\mu+\frac12}(z) H^{(2)}_{-i\mu-\frac12}(z)\Big)
\nn
\\
&  ~~~~  \times \sin k |\vec{x}-\vec{x}'| d k.\label{equi}
\el
By the recurrence relations of the Hankel functions,   \eqref{equi} can be written as
\bl
\langle 0| \bar\psi(x)\psi(x') |0 \rangle
 & =i\frac{H^3}{4\pi\sigma}\int_0^\infty
\frac{d}{dz}   \Big(H^{(1)}_{\nu}(z)H^{(2)}_{\nu}(z)  \Big)
z^2\sin (\sigma z) d z
\nn \\
& +  \frac{H^3}{4\pi\sigma}(i-2\mu)\int_0^\infty
H^{(1)}_{\nu}(z)H^{(2)}_{\nu}(z)
z\sin (\sigma z) d z,\label{traceofGreenfunctiion}
\el
where  $\sigma \equiv aH|\vec{x}-\vec{x}'|$,  and  $\nu\equiv-\frac12-i\mu$.
We now calculate the first integral  in \eqref{traceofGreenfunctiion}.
\bl
Int_1 &\equiv \int_0^\infty
\frac{d}{dz}  \Big(H^{(1)}_{\nu}(z)H^{(2)}_{\nu}(z)  \Big)
z^2\sin (\sigma z) d z.
\label{int1}
\el
By use  of  the following formulae  (see  Ref. \cite{Olver2010} and (6.671.5) in Ref. \cite{SUM1980})
\bl
J_{\nu}(z)^2+Y_{\nu}(z)^2
& =\frac{8}{\pi^2}\int_{0}^{\infty}\cosh(2\nu t)K_{0}(2 z \sinh t) \,  dt,
\label{JY}
\\
\frac{ d K_{0}(x)}{dx }  & =-K_{1}(x),
\\
\int_{0}^{\infty}K_{\nu}(a z)
\sin (b z) d z  & = \frac{1}{4}\pi a^{-\nu} \csc \big(\frac{\nu\pi}{2} \big)
\frac{1}{\sqrt{a^2+b^2}}\big( \big[(b^2+a^2)^{\frac12}+b\big]^{\nu}
-\big[(b^2+a^2)^{\frac12}-b\big]^{\nu}\big) ,
\el
the integration \eqref{int1}  can be written as
\bl
Int_1 =-\frac{8}{\pi^2}(-\frac{\partial^2}{\partial \sigma^2})
\int_0^\infty2\sinh t\cosh(2\nu t)
\Big( \frac{\pi}{4} (2\sinh t)^{-1}
      \frac{2\sigma}{\sqrt{4\sinh t^2+\sigma^2}} \Big) dt       .
\el
Further simplification gives
\bl
Int_1 &=-\frac{12\sigma}{\pi}\int_0^\infty
\frac{4\sinh^2 t\cosh(2\nu t)}{(4\sinh^2 t+\sigma^2)^{\frac52}}dt
\nn
\\
&=3\times2^{-\frac12}\frac{\sigma}{\pi}\int_0^\infty
\frac{(\cosh(\nu T)-2^{-1}\cosh ((1+\nu) T))-2^{-1}\cosh ((\nu-1) T))}{((\frac{\sigma^2}{2}-1)+\cosh T)^{\frac52}}dT,\label{int1changev}
\el
where the integration variable  $T  \equiv 2 t $.
Using  relations (14.3.15), (15.1.1) and (15.1.2)  in Ref. \cite{Olver2010}
 in \eqref{int1changev} leads to
\bl
Int_1
&=\frac{\sigma}{4\pi}\Gamma(\nu+\frac32)\Gamma(\frac32-\nu)\Big(
(\frac94-\nu^2)
\,_2F_1(\frac{5}{2}-\nu,\frac{5}{2}+\nu;3;1-\frac{\sigma^2}{4})\nn
\\
&~~~~~~~~   -\frac12(\nu+\frac{5}{2})(\nu+\frac{3}{2})
\,_2F_1(\frac32-\nu,\frac72+\nu;3;1-\frac{\sigma^2}{4})\nn
\\
&~~~~~~~~   -\frac12
(\frac{5}{2}-\nu)(\frac{3}{2}-\nu)
\,_2F_1(\frac72-\nu,\nu+\frac32;3;1-\frac{\sigma^2}{4})\Big),\label{firstinteGren}
\el
where $\Gamma(x+1)=x\Gamma(x)$ has been used
and  $\,_2F_1(a,b;c;d)$ is the hypergeometric function.

Now calculate  the second integration in \eqref{traceofGreenfunctiion}
\bl
Int_2   \equiv  \int_0^\infty H^{(1)}_{\nu}(z)H^{(2)}_{\nu}(z)
z\sin (\sigma z) d z.
\label{int2}
\el
Using  \eqref{JY}  and  the following  formula \cite{SUM1980}
\bl
\int_{0}^{\infty}K_{0}(\beta z)
\cos (\alpha z) d z=\frac{\pi}{2\sqrt{\alpha^2+\beta^2}} ,
\el
 the integration \eqref{int2} can be expressed  as
\bl
Int_2 =  \frac{8}{\pi^2}(-\frac{\partial}{\partial \sigma})\int_0^\infty\cosh(2\nu t)
\frac{\pi}{2\sqrt{4\sinh^2t+\sigma^2}}dt
 ,  \label{nfre}
\el
which  is written as
\bl
Int_2
&=\frac{2^{-\frac12}\sigma}{\pi}\int_0^\infty
\frac{\cosh(\nu T)}{((\frac{\sigma^2}{2}-1)+\cosh T)^{\frac32}}d T,\label{dheuwih}
\el
where the integration variable $t$ has changed from $t$ to $T/2$.
Using the relations (14.12.4), (14.3.15), (15.1.1), and (15.1.2) in Ref. \cite{Olver2010} in
\eqref{dheuwih} yields
\bl
Int_2
&=\frac{\sigma}{2\pi}
\Gamma(\nu+\frac32)\Gamma(\frac32-\nu)
 \Gamma(2)^{-1}\,_2F_1(\frac32-\nu,\nu+\frac32;2;1-\frac{\sigma^2}{4}).\label{secondinteGren}
\el
Plugging \eqref{firstinteGren} and \eqref{secondinteGren} into \eqref{traceofGreenfunctiion} yields the equal-time correlation  function
\bl
&\langle0|\bar\psi(\vec{x},t)\psi(\vec{x'},t)|0\rangle\nn
\\
= & \frac{H^3}{\pi^2}\Gamma(\nu+\frac32)\Gamma(\frac32-\nu)\Big(
-\frac{i}4\nu
\,_2F_1(\frac32-\nu,\frac32+\nu;2;1-\frac{\sigma^2}{4})
\nn \\
& +\frac{i}{16}(\frac94-\nu^2)
\,_2F_1(\frac{5}{2}-\nu,\frac{5}{2}+\nu;3;1-\frac{\sigma^2}{4})\nn
\\
&
-\frac{i}{32}(\nu+\frac{5}{2})(\nu+\frac{3}{2})
\,_2F_1(\frac32-\nu,\frac72+\nu;3;1-\frac{\sigma^2}{4})
\nn \\
& -\frac{i}{32} (\frac{5}{2}-\nu)(\frac{3}{2}-\nu)
\,_2F_1(\frac72-\nu,\nu+\frac32;3;1-\frac{\sigma^2}{4})\Big)
   .\label{notgeneral}
\el
By the maximal symmetry of de Sitter space,
the correlation   function depends in general on the one-half of the squared geodesic interval $\sigma_2$ of \eqref{sigma2},
 so we can replace
\bl
&-\frac12\sigma^2\rightarrow\sigma_2, \label{replacement}
\el
in \eqref{notgeneral} to give
the non-equal time  correlation   function \eqref{unrcorr}.

In the massless limit $\mu =0$, $\nu = -\frac12 $, the correlation function
\eqref{notgeneral} reduces to zero.

\end{document}